\documentclass[aoas]{imsart}

\RequirePackage{amsthm,amsmath,amsfonts,amssymb}
\RequirePackage[authoryear]{natbib}
\RequirePackage[colorlinks,citecolor=blue,urlcolor=blue]{hyperref}
\RequirePackage{graphicx}
\usepackage{algorithm}
\usepackage{algorithmic}
\usepackage{makecell}
\usepackage{subfigure}
\usepackage{booktabs}
\usepackage{xcolor}

\def\n{\nonumber}

\def\beq{\begin{equation}}
\def\eeq{\end{equation}}
\def\beqr{\begin{eqnarray}}
\def\eeqr{\end{eqnarray}}
\def\beqrs{\begin{eqnarray*}}
\def\eeqrs{\end{eqnarray*}}
\def\bet{\begin{theorem}}
\def\eet{\end{theorem}}
\def\bel{\begin{lemma}}
\def\eel{\end{lemma}}
\def\bep{\begin{proposition}}
\def\eep{\end{proposition}}
\def\bg{\begin{figure}[tbph]\begin{center}}
\def\eg{\end{center}\end{figure}}

\def\bc{\begin{center}}
\def\ec{\end{center}}

\def\vare{\varepsilon}

\def\wh{\widehat}

\def\SE{\widehat{\mbox{SE}}}

\def\b0{\mathbf{0}}

\def\mR{\mathbb{R}}

\def\mL{\mathcal L}

\def\mF{\mathcal F}

\def\mV{\mathbf{V}}
\def\mX{\mathbf{X}}
\def\mW{\mathbf{W}}

\def\mY{\mathbf{Y}}

\def\var{\mbox{var}}

\def\tr{\mbox{tr}}
\def\argmin{\mbox{argmin}}

\def\diag{\mbox{diag}}

\def\CLS{\rm CLS}
\def\CLE{\rm CLE}
\def\LS{\rm LS}

\newcommand{\mE}{{\mathbf E}}
\newcommand{\mcE}{{\mathcal E}}

\startlocaldefs
\theoremstyle{plain}

\newtheorem{theorem}{Theorem}[section]
\newtheorem{proposition}[theorem]{Proposition}
\newtheorem{corollary}[theorem]{Corollary}
\newtheorem{lemma}[theorem]{Lemma}
\theoremstyle{remark}

\newtheorem{remark}{Remark}

\endlocaldefs

\begin{document}

\begin{frontmatter}
\title{Bias-Correction for Privacy-Protected Spatial Autoregressive Models with  Application \\to Restaurant Network Analysis}
\runtitle{Privacy-Protected Spatial Autoregressive Models}

\begin{aug}

\author[A]{\fnms{Danyang}~\snm{Huang} \ead[label=e1]{dyhuang89@126.com}},
\author[A]{\fnms{Ziyi}~\snm{Kong}\ead[label=e2]{kzymiao@163.com}},
\author[B]{\fnms{Shuyuan}~\snm{Wu}\thanks{Corresponding author.
}\ead[label=e3]{wushuyuan@mail.sufe.edu.cn}}
\and
\author[C]{\fnms{Hansheng}~\snm{Wang}\ead[label=e4]{hansheng@pku.edu.cn}}
\address[A]{Center for Applied Statistics and School of Statistics, Renmin University of China\printead[presep={ ,\ }]{e1,e2}}

\address[B]{School of Statistics and Data Science, Shanghai University of Finance and Economics \printead[presep={,\ }]{e3}}
\address[C]{Guanghua School of Management, Peking University \printead[presep={,\ }]{e4}}
\end{aug}

\begin{abstract}
Spatial autoregressive (SAR) models and their extensions are important tools for studying network effects. However, with an increasing emphasis on data privacy, data providers often implement protection measures that render standard SAR models inapplicable. In this study, we introduce a privacy-protected SAR model that incorporates noise into both the response and covariates to meet privacy requirements.
With noise present in both components, the traditional quasi-maximum likelihood estimator becomes difficult to compute because the likelihood function cannot be directly formulated.
To bypass this hurdle, we begin with a pseudo-likelihood approach, initially omitting the noise in the covariates. A Newton-Raphson algorithm is then applied to compute the estimator; however, the estimator is biased. To address this, we propose a bias-corrected  Newton-Raphson-type algorithm that simultaneously accounts for noise in both the response and covariates. We further show, under appropriate regularity conditions, that the resulting estimator is consistent and asymptotically normal.
To further enhance computational efficiency, we also develop a bias-corrected least squares estimator. Several extensions are discussed, and the finite-sample performance of the proposed methods is evaluated through extensive simulations.
We apply the proposed methodology to restaurant transaction data from a third-party payment platform. Our method identifies a statistically significant competitive network effect among restaurants and further reveals meaningful restaurant-customer interaction patterns.
\end{abstract}

\begin{keyword}
\kwd{Spatial Autoregressive Model}
\kwd{Privacy Protection}
\kwd{Bias-Corrected Estimation}
\kwd{Least Squares Estimation}
\end{keyword}

\end{frontmatter}

\section{Introduction}
\subsection{Motivating Application and Problem Setup}
In modern data science applications, network data—comprising nodes and their observed relationships—are widely used to model complex interactions among entities. Such data naturally induce dependence between observations, as nodes are connected through edges that capture social, spatial, or behavioral relationships \citep{lee2010estimation,hardle2016tenet,zhu2020multivariate}. This network-induced dependence is a central feature in many fields, including business operations, healthcare systems, and consumer behavior analysis. However, a growing number of applications involve sensitive data, where either node-level variables or the network structure itself contain private information. Protecting such information has become a prerequisite for responsible data analysis.

This study is motivated by a collaborative project with one of the largest aggregate payment platforms in China. The objective is to investigate how network effects and merchant-level characteristics influence restaurant transaction performance and how the merchants and customers interact. The underlying dataset includes sensitive business information such as transaction volume, average transaction value, and customer loyalty rates, as well as network relationships inferred from customer mobility patterns. Disclosing the nodal variables such as response or covariates could expose confidential operational information or enable unfair competition. Therefore, any analysis must respect strict privacy constraints, requiring methods that protect the nodal attributes.

The spatial autoregressive (SAR) model has proven to be a powerful tool for modeling network dependence \citep{ord1975estimation,anselin2001,Lee:2004}. However, most existing SAR studies assume fully observable data and do not account for privacy concerns. In practice, however, privacy protection imposes substantial constraints on data accessibility and usability. 
These challenges highlight the urgent need for a statistically valid and computationally feasible framework that enables network-based inference under privacy constraints. To this end, developing a general privacy-protected methodology for the SAR family of models is both necessary and timely.

\subsection{Related Literature}
 The SAR model assumes that the response at each node depends linearly on the responses of its neighbors. This formulation captures network-induced dependence in a parsimonious and interpretable manner, making it a popular choice in many applied fields, including economics, marketing, and urban analytics \citep{anselin2008spatial_typical, chen2013impact, zhu2017network, beenstock2019econometric}. 
The classical SAR models have inspired numerous follow-up studies. For example, \cite{lee2010estimation} introduced the SAR panel model with individual effects and SAR disturbances. \cite{yang2016} and \cite{zhu2020multivariate} developed a multivariate SAR model that models multivariate responses collected from each node. Various partially linear semiparametric spatial models were also developed to allow the SAR parameter to meaningfully vary across different nodes \citep{Su2012,Malikov2017}. The spatial dynamic panel data models were proposed to model time and network dependence simultaneously \citep{yu2008quasi,lee2014efficient,li2017fixed}. To compute the maximum likelihood estimator for various SAR models, the determinant of a large-scale matrix must be computed. This creates a significantly high level of computational complexity. To alleviate the computational cost, various efficient computing algorithms were developed \citep{pace2000method,lesage2007matrix,zhou2017}. In particular, the least squares method of \cite{huang2019} and \cite{zhu2020multivariate} has been demonstrated to be practically useful.

Regarding privacy protection measures, classical methods include, but are not limited to, data swapping \citep{Reiter2005}, data imputation \citep{Raghunathan2003,Reiter2005}, posterior distribution sampling \citep{Hu2018,Wilde2021}, and noise addition \citep{Dwork2006,Wilde2021,Ito2021}. Among these measures, noise addition has arguably gained great popularity due to both its implementational simplicity and tunable privacy guarantees through noise-level specification. Modern implementations of noise addition are formally grounded in differential privacy theory \citep{Dwork2006,Dwork2014}, with two dominant paradigms. One is differential privacy for data 
\citep{kasiviswanathan2011can,duchi2013local,joseph2019role,wang2020comprehensive}, 
it applies protections directly to individual records, typically through local mechanisms like randomized response \citep{warner1965randomized,kairouz2014extremal} or value perturbation \citep{nguyen2016collecting,duchi2018minimax}. The other one, differential privacy for model, instead protects aggregated information by adding noise during model training or inference, as seen in techniques like differentially private SGD \citep{abadi2016deep,bu2020deep}, noisy gradient descent \citep{cai2021cost,avella2023differentially}, noisy Newton \citep{avella2023differentially} and peeling method for high-dimensional data \citep{dwork2018differentially,cai2021cost}. Besides, privacy-aware frameworks like federated learning have emerged, and privacy is preserved not through noise addition but via decentralized model updates and restricted data sharing \citep{mcmahan2017communication,bagdasaryan2020backdoor}. 

With regard to network analysis, existing differential privacy approaches generally fall into two distinct categories of privacy protection \citep{chen2023private}. The first category is edge-DP, which considers networks differing by a single edge \citep{karwa2016inference,wang2022two,chang2024edge}. And the second one is node-DP, a stronger requirement that allows two networks to differ by a single node, including all edges incident to that node \citep{chen2013recursive, day2016publishing}. However, adapting this approach to privacy-protected spatial data introduces nontrivial difficulties. When data companies share data, they aim to obtain accurate estimates while also protecting sensitive variables, whether the covariates \(X\) or the response \(Y\). This necessitates adding noise to both \(X\) and \(Y\). 


This issue is related to measurement errors in variables in traditional statistical research, where measurement errors in covariates have received considerable attention \citep{Fuller1987,Stefanski1987,Staudenmayer2005,Carroll2006,Buonaccorsi2010,Hausman2021}. Tools to deal with measurement error include, but are not limited to, simulation-extrapolation \citep{cook1994simulation, Novick2002, bertrand2017inference}, instrumental variables \citep{schennach2007instrumental,gustafson2007measurement,jiang2020measurement,Luo2022}, and the corrected score method \citep{Stefanski1987,Nakamura1990,tsiatis2004locally,wang2012corrected}. In the analysis of SAR models, scholars have also focused on addressing covariate measurement error. For example, \cite{Li2009} adopted structural model-based approaches to integrate out \(X\) when the distribution of \(X\)'s measurement error is known. \cite{Luo2022} proposed the 3SLS method which uses instrumental variables to deal with measurement error in \(X\).
However, current literature on addressing measurement error in SAR models rarely considers scenarios where both \(X\) and \(Y\) contain noise.
In linear models, measurement errors in the continuous variable \(Y\) do not affect the unbiasedness of the parameter estimates \citep{Hausman2001}. However, in SAR models, adding artificial noise to \(Y\) inevitably leads to biased parameter estimates. Furthermore, in data collaborations, the sample size is typically large. However, existing literature on SAR with measurement errors seldom considers the computational complexity. Therefore, developing accurate and computationally efficient parameter estimation methods for SAR models for both noisy \(X\) and \(Y\) is of great importance in real practice.

\subsection{Our Contribution}
In this study, we propose a privacy-protected SAR (PSAR) model to meet the privacy-protection requirements. To obtain more accurate estimates, we assume that the data company has predetermined and disclosed the level of noise. Unlike previous SAR models, the unavailability of true values for both the response and some explanatory variables makes the classical quasi-maximum likelihood estimator (QMLE) inapplicable. To address this challenge, inspired by the corrected score method \citep{Stefanski1987,Nakamura1990}, we first derive the likelihood function based on the true explanatory variables and observed responses. This approach allows for an explicit expression of the log-likelihood function and the derivation of its first- and second-order derivatives, which are, however, biased. Nonetheless, the bias can be analytically expressed under appropriate regularity conditions and subsequently corrected. Subsequently, a Newton-Raphson-type algorithm can be designed to obtain a corrected likelihood estimator (CLE). Nonetheless, computing the CLE becomes highly challenging for large-scale networks due to the computational complexity associated with high-dimensional matrix operations, such as determinant and inverse computations. To address this issue, we propose a corrected least squares estimator (CLS) for the PSAR model inspired by \cite{meinshausen2006variable}, \cite{huang2019}, and \cite{zhu2020multivariate}, and the concept of bias correction. In the theoretical analysis, we establish the identifiability, asymptotic normality, and numerical convergence properties for the estimators. 
We then extend the two privacy-protected methods beyond the classical SAR model to a more flexible multi-mode, multivariate SAR framework, incorporating perturbations to the network structure $W$  to enhance privacy protection. 
It should be noted that the data perturbation for privacy protection is imposed by the data provider at the data-release stage, while our methodological contribution focuses on bias-corrected estimation and inference for SAR models based on the released perturbed data.
 Applying our methodology to restaurant network data from a third-party payment company, we identify a significant competitive effect among restaurants and meaningful restaurant-customer interaction patterns under privacy protection.

The remainder of this paper is organized as follows. Section 2 describes the estimation methods in detail. Section 3 presents the theoretical properties from both statistical and numerical perspectives and provides a discussion of generalization for these methods. Section 4 presents the simulation studies, then Section 5 shows a real data example. Section 6 presents the concluding remarks. All technical details are relegated to the supplementary material. 

\section{Privacy-protected spatial autoregressive model and estimators}
\subsection{Model and notations}

Consider a network with a total of $N$ nodes. To describe the network structure (directed or undirected), we define an adjacency matrix $A=(a_{i_1i_2})\in \{0,1\}^{N\times N}$ ($1\leq i_1,i_2\leq N$). Each element $a_{i_1i_2}=1$ if there exists an edge from node $i_1$ to $i_2$ ($i_1\neq i_2$); otherwise, $a_{i_1i_2}=0$. We assume $a_{ii}=0$ for $1\leq i\leq N$. We define  the weighting matrix $W =(w_{ij}) \in \mathbb{R}^{N \times N}$ with element $w_{ij} = a_{ij}/d_i$, where $d_i =\sum_{j=1}^N a_{ij} $ represents the nodal out-degree of node $i$. Throughout the remainder of this paper, we assume $d_i>0$ for every $1\leq i\leq N$. Otherwise, nodes with $d_i=0$ should not be included for analysis.

Next, let $Y_{i}$ be the response collected from the $i$-th node ($1\leq i\leq N$) and $\mathbf{x}_{i}\in\mathbb{R}^p$ be the associated covariate vector. Write $\mathbf{Y} = ({Y}_{1}, \cdots,{Y}_{N})^\top\in \mathbb{R}^{N}$ as the response vector, $\mathbf{X}=(\mathbf{x}_{1},\cdots,\mathbf{x}_{N})^\top\in \mathbb{R}^{N\times p}$ as the associated covariate matrix representing all explanatory variables, and $\mathbf{E}=(e_{1}, \cdots,e_{N})^\top\in \mathbb{R}^{N}$ as the error vector, where $e_{i}$s are independent and identically distributed with mean 0 and covariance $\sigma^2_0$.  If all the response and explanatory variables can be faithfully observed, we consider the following SAR model \citep{ord1975estimation,anselin2008spatial_typical},
\begin{equation}
\begin{aligned}
\label{sarp}
\mathbf{Y}= \rho_0 W{\mathbf{Y}} + \mathbf{X}\beta_0+\mathbf{E}{,}
\end{aligned}
\end{equation}
where $\rho_0 \in \mathbb{R}$ is the network autocorrelation parameter measuring the network effect, and $\beta_0 \in \mathbb{R}^p$ is the regression-coefficient vector reflecting the effect of nodal covariates. 
Define \( S_0 = I_N - \rho_0 W \) as the true model matrix, where \( \rho_0 \) is the true parameter and \( I_N \in \mathbb{R}^{N \times N} \) is the identity matrix. We later use \( S = I_N - \rho W \) to denote a matrix corresponding to an arbitrary \( \rho \) in the parameter space. We then have $\mathbf{Y} = S_0^{-1}(\mathbf{X}\beta_0+\mathbf{E})$. By omitting the constant term, the log-likelihood function for the parameter of interest $\theta=(\rho,\beta^\top,\sigma^2)^\top\in\mathbb{R}^{p+2}$ can be readily derived.
Accordingly, a QMLE  for $\theta$ can be obtained. 
Its asymptotic properties have been well studied. For example, refer to \cite{Lee:2004} and \cite{anselin2013spatial}. It is remarkable that the SAR models could be generalized to allow for different type of nodes with multiple responses. For intuitively illustrate the estimation procedure, we consider privacy protection for the classical SAR model first and then consider the generalization of the SAR models.

In privacy-protection scenario, we assume that the response $\mathbf{Y}$ and part of the explanatory covariates cannot be faithfully observed. Specifically, for the covariates, we assume that $\mathbf{X}=(\mathbf{X}_1,\mathbf{X}_2)$ with $\mathbf{X}_1\in\mathbb{R}^{N\times p_1}$, $\mathbf{X}_2\in\mathbb{R}^{N\times p_2}$, and $p_1+p_2=p$. Here, we define $\mathbf{X}_1$ to be the matrix collecting the faithfully observed covariates, and $ \mathbf{X}_2^*=\mathbf{X}_2+\mathcal{E}_x\in\mathbb{R}^{N\times p_2}$ to be the matrix collecting explanatory variables with artificially created error matrix $ \mathcal{E}_x=(\varepsilon_{x,ij})\in\mathbb{R}^{N\times p_2}$. Further, we assume that each element $ \varepsilon_{x,ij}$ ($ 1\leq i\leq N,1 \leq j\leq p_2$) in $\mathcal{E}_x$ is independently and identically distributed with mean 0 and predefined variance $\lambda_x^2$.
Moreover, for the response, a data user can only observe $\mathbf{Y}^*=\mathbf{Y}+\mathcal{E}$ with an artificially created noise vector $\mathcal{E}=(\varepsilon_{1},\cdots,\varepsilon_{N})^\top\in\mathbb{R}^N$. Here, we assume that different $\varepsilon_{i}$s are independent and identically distributed with mean 0 and known variance $\lambda^2$.
We assume here that data users cannot observe the true response variables $\mathbf{Y}$ and some explanatory variables $\mathbf{X}_2$. However, to better utilize the data, they are aware of the privacy-protection parameters $\lambda^2$ and $\lambda_x^2$. This assumption ensures the privacy of the data while simultaneously facilitating better estimation results for data users. Notably, if \(\lambda^2\) is unknown, an identification issue arises for the parameter when \(\rho = 0\). However, since \(\lambda^2\) is assumed to be known, this identification problem is eliminated. Following \citet{joseph2019role} and \citet{wang2020comprehensive}, we can establish a strict local differential privacy guarantee for the above data release mechanism; see Appendix C.6 for details. 

We next study the statistical properties of the resulting estimator. To this end,  
we can derive the expression for $\mathbf{Y}^*$ based on the observed $\mathbf{X}^*=(\mathbf{X}_1,\mathbf{X}^*_2)$ as,
\begin{equation}\label{sar_dp_ge}
\begin{aligned}
\mathbf{Y}^* = S_0^{-1}\Big\{\mathbf{X}_1\beta_{01}+(\mathbf{X}_2^*-\mathcal{E}_x)\beta_{02}+\mathbf{E}\Big\}+\mathcal{E},
\end{aligned}
\end{equation}
where $\beta_{01}\in\mathbb{R}^{p_1}$ represents the coefficient corresponding to the truly observed covariates, $\beta_{02}\in\mathbb{R}^{p_2}$ is the coefficient corresponding to those with noise added, and we have $\beta_0=(\beta_{01}^\top,\beta_{02}^\top)^\top$. 

For convenience, we refer to this model in \eqref{sar_dp_ge} as the privacy-protected spatial autoregressive (PSAR) model. Notably, model \eqref{sar_dp_ge} only represents the relationship between the observable $\mathbf{Y}^*$ and $\mathbf{X}^*$. However, the challenge here is that $\mathbf{Y}^*$ is generated by true $\mathbf{X}$, which is not fully observable to data users.
\subsection{Corrected likelihood estimator}

To accomplish model estimation, the foremost method worth considering is QMLE. However, note that the likelihood function for \eqref{sar_dp_ge} is hard to be spelled out because $\mY^*$ is generated according to the true $\mX$. 
As $\mathbf{X}^*$ is noise-added, the conditional likelihood of $\mathbf{Y}^*$ given $\mathbf{X}$ is no longer
available, we need to rely on the joint likelihood of $\mathbf{Y}^*$ and $\mathbf{X}^*$. Unfortunately, this likelihood is intractable since an $N$-dimensional integration with respect to the noise added to $\mathbf{X}$ is inevitably involved. 
To address this issue, we begin the analysis by assuming that $\mX_2$ is known, and we consider the underlying generating process of the observed $\mY^*$. Therefore, we have
$
\mY^* = S_0^{-1}(\mathbf{X}\beta_0+\mathbf{E})+\mathcal{E}.
$
The negative log-likelihood function for  $\theta$  based on $\mY^*$ and $\mX$ is,
\beqr
\label{eq:lik}
\mL(\theta)=\mL(\theta,\mY^*,\mX)=-\log|S|+\frac{1}{2}\log|\Omega|+\frac{1}{2}\Big(S\mY^*-\mX\beta\Big)^\top\Omega^{-1}\Big(S\mY^*-\mX\beta\Big),
\eeqr
where $\Omega=\sigma^2 I_N+\lambda^2SS^{\top}$. Ideally, we can apply the classical Newton-Raphson algorithm to solve the optimization problem $\arg\min_\theta \mL(\theta)$, and obtain estimates with desirable properties. Unfortunately, equation (\ref{eq:lik}) is computationally infeasible in real practice because  we can only observe $\mX_2^*$.
Direct substitution of $\mX_2$ with $\mX_2^*$ in (\ref{eq:lik}) to obtain $\mL^*(\theta)=\mL(\theta,\mY^*,\mX^*)$ for feasible computation will inevitably result in biased estimation. This is because the key statistical estimators derived from $\mL^*(\theta)$, including the first- and second-order derivatives, are all biased.

To clarify this point, we begin by analyzing the difference between the first-order derivatives of $\mL(\theta)$ and $\mL^*(\theta)$. Define $ \dot\mL(\theta)$ and $ \dot\mL^*(\theta)$ to be the first-order derivative of $\mL(\cdot)$ and $\mL^*(\cdot)$  with respect to $\theta$, respectively.
Then, we can calculate that $E\big\{ \dot{\mL}(\theta_0)  \big\}=\mathbf{0}_{p+2}$ and,
\beqr
E\big\{\dot\mL^{*\rho}(\theta_0)\big\}&=&\lambda^2\lambda_x^2\beta_{02}^\top\beta_{02}\tr\Big(\Omega_0^{-1}WS_0^\top\Omega_0^{-1}\Big),\label{eq:CLEfrho}\\
E\big\{\dot\mL^{*\beta}(\theta_0)\big\}&=&\Big(\mathbf{0}_{p_1}^\top,\lambda_x^2\tr(\Omega_0^{-1})\beta_{02}^\top\Big)^\top,\label{eq:CLEfbeta}\\
E\big\{\dot\mL^{*\sigma^2}(\theta_0)\big\}&=&- \frac{1}{2}\lambda_x^2\beta_{02}^\top\beta_{02}\tr(\Omega_0^{-2})\label{eq:CLEfsigma},
\eeqr
\noindent
where 
$\dot\mL^{*\rho}(\theta)$, $\dot\mL^{*\beta}(\theta)$, and $\dot\mL^{*\sigma^2}(\theta)$ are the first-order derivatives of $\mL^*(\theta)$ with respect to $\rho$, $\beta$, and $\sigma^2$, respectively. See Appendix A.1 for detailed verifications. We define this difference for the first-order derivative caused by  $\mX^*$ as $\Delta\mathbf{S}_{\rm CL}(\theta_0)=E\big\{\dot\mL^*(\theta_0)\big\}-E\big\{\dot\mL(\theta_0)\big\}$.

Similarly, define the second-order derivative for $\mL(\theta)$ and $\mL^*(\theta)$ as $\ddot\mL(\theta)$ and $\ddot\mL^*(\theta)$, respectively. Define $\mW_S=WS^{\top}+SW^{\top}$.
In this way, the difference between the expectations of second-order derivatives for $\mL^*(\theta_0)$ and $\mL(\theta_0)$ could be calculated and defined as,
\beqr\label{eq:CLEs}
\Delta\mathbf{H}_{\rm CL}(\theta_0)= E\big\{\ddot\mL^*(\theta_0)\big\}-E\big\{\ddot\mL(\theta_0)\big\}=\left(
\begin{array}{ccc}
\mathbf{H}_{\rm CL}^{\rho\rho} & \mathbf{H}_{\rm CL}^{\rho\beta\top} & \mathbf{H}_{\rm CL}^{\rho\sigma} \\
\mathbf{H}_{\rm CL}^{\rho\beta} & \mathbf{H}_{\rm CL}^{\beta\beta} & \mathbf{H}_{\rm CL}^{\beta\sigma} \\
\mathbf{H}_{\rm CL}^{\rho\sigma} & \mathbf{H}_{\rm CL}^{\beta\sigma\top} & \mathbf{H}_{\rm CL}^{\sigma\sigma} \\
\end{array}
\right).
\eeqr
It could be verified that $\Delta\mathbf{H}_{\rm CL}(\theta_0)$ is symmetric and
\begin{equation*}
\begin{aligned}
\mathbf{H}_{\rm CL}^{\rho\rho}=&\lambda^4\lambda_x^2\beta_{02}^\top\beta_{02}\tr(\Omega_0^{-2}\mW_{S_0}\Omega_0^{-1}\mW_{S_0})-\lambda^2\lambda_x^2\beta_{02}^\top\beta_{02}\tr(\Omega_0^{-2}WW^\top),\\
\mathbf{H}_{\rm CL}^{\rho\beta}=&\Big(\mathbf{0}_{p_1}^\top,\lambda^2\lambda_x^2\tr(\Omega_0^{-2}\mW_{S_0})\beta_{02}^\top\Big)^\top, ~~~~\mathbf{H}_{\rm CL}^{\rho\sigma}=-\lambda^2\lambda_x^2\beta_{02}^\top\beta_{02}\tr\Big(\Omega^{-3}_0\mW_{S_0}\Big),\\
\mathbf{H}_{\rm CL}^{\beta\beta}=&\Big(\mathbf{0}_{p_1\times p_1},\mathbf{0}_{p_1\times p_2};\mathbf{0}_{p_2\times p_1},\lambda_x^2\tr(\Omega_0^{-1})I_{p_2}\Big),\\
\mathbf{H}_{\rm CL}^{\beta\sigma}=&\Big({\bf 0}_{p_1}^\top,-\lambda_x^2\tr(\Omega_0^{-2})\beta_{02}^\top\Big)^\top,~~~\mathbf{H}_{\rm CL}^{\sigma\sigma}=\lambda_x^2\beta_{02}^\top\beta_{02}\tr(\Omega_0^{-3}).\\
\end{aligned}
\end{equation*}

\noindent See Appendix A.1 for detailed verifications.
Based on the explicitly expressed bias in the derivatives led by $\mX_2^*$, a natural method to estimate parameter $\theta_0$ is to reconstruct an estimator using the bias-corrected first- and second-order derivatives. We refer to this estimator as {\it the corrected likelihood estimator} (CLE) , which is denoted as $\hat\theta_{\rm CLE}$. Accordingly, the estimator $\hat\theta_{\rm CLE}$ is defined as the solution to $\dot\mL^*(\theta)-\Delta\mathbf{S}_{\rm CL}(\theta)=\mathbf{0}_{p+2}.$ Accordingly, an iterative algorithm can be developed. At each iteration, we conduct three steps: (1) \textbf{(Calculating)} compute the first- and second-order derivatives based on the observed $\mX^*$ and $\mY^*$; (2) \textbf{(Debiasing)} perform bias correction for the derivatives; and (3) \textbf{(Updating)} conduct a Newton-Raphson-type iterative formula based on the corrected derivatives.

To be more specific,
let $\hat\theta_{\rm CLE}^{(0)}=(\hat\rho_{\rm CLE}^{(0)},\hat\beta_{\rm CLE}^{(0)\top},\hat\sigma^{2(0)}_{\rm CLE})^\top\in\mR^{p+2}$ be the initial estimator, which could be, for example, the QMLE estimator assuming that $\mX_2^*$ as the true $\mX_2$.  Let $\hat\theta_{\rm CLE}^{(t)}=(\hat\rho_{\rm CLE}^{(t)},\hat\beta_{\rm CLE}^{(t)\top},\hat\sigma^{2(t)}_{\rm CLE})^\top\in\mR^{p+2}$ be the estimator obtained  in the $t$-th iteration, and $\hat S^{(t)}$, $\hat\Omega^{(t)}$ be $S$, $\Omega$ with the plugged-in estimator $\hat\theta_{\rm CLE}^{(t)}$.
We next consider how to update $\hat\theta_{\rm CLE}^{(t)}$ to $\hat\theta_{\rm CLE}^{(t+1)}$.
We start from the parameter $\gamma_0=(\rho_0,\beta_0^\top)^\top\in\mR^{p+1}$ and its update $\hat\gamma_{\rm CLE}^{(t+1)}=(\hat\rho_{\rm CLE}^{(t+1)},\hat\beta_{\rm CLE}^{(t+1)\top})^\top$.
It could be calculated that, for the first-order derivative, $\Delta\textbf{S}_{\rm CL}^{(t)}
=\Big(\lambda^2\lambda_x^2\hat\beta_{2}^{(t)\top}\hat\beta_{2}^{(t)}\tr\Big\{(\hat\Omega^{(t)})^{-1}W\hat S^{(t)\top}(\hat\Omega^{(t)})^{-1}\Big\},
{\bf0}_{p_1}^\top, \lambda_x^2\tr\big\{(\hat\Omega^{(t)})^{-1}\big\} \hat\beta^{(t)\top}_{2}\Big)\in\mR^{p+1}$. Similarly, define $\Delta\textbf{H}_{\rm CL}^{(t)}
=\Big(\mathbf{H}_{\rm CL}^{\rho\rho(t)} ,\mathbf{H}_{\rm CL}^{\rho\beta(t)\top};\mathbf{H}_{\rm CL}^{\rho\beta(t)},\mathbf{H}_{\rm CL}^{\beta\beta(t)}  \Big)\in\mR^{(p+1)\times(p+1)}
$, which could be calculated with plugged in $\hat\theta_{\rm CLE}^{(t)}$ correspondingly. Then, $\hat\gamma_{\rm CLE}^{(t+1)}$ can be obtained using a corrected Newton-Raphson-type algorithm as,
\beqrs
\hat\gamma_{\rm CLE}^{(t+1)}=\hat\gamma_{\rm CLE}^{(t)}- \Big\{\ddot\mL^{*\gamma}(\hat\theta_{\rm CLE}^{(t)})- \Delta \textbf{H}_{\rm CL}^{(t)} \Big\}^{-1}\Big\{\dot\mL^{*\gamma}(\hat\theta_{\rm CLE}^{(t)}) - \Delta \textbf{S}_{\rm CL}^{(t)}  \Big\} \label{eq:CLErho},
\eeqrs
where $\dot\mL^{*\gamma}(\hat\theta_{\rm CLE}^{(t)})=\Big(\dot\mL^{*\rho}(\hat\theta_{\rm CLE}^{(t)}),\dot\mL^{*\beta\top}(\hat\theta_{\rm CLE}^{(t)})\Big)^\top$ and  $\ddot\mL^{*\gamma}(\hat\theta_{\rm CLE}^{(t)})=\Big(\ddot\mL^{*\rho\rho}(\hat\theta_{\rm CLE}^{(t)}),\ddot\mL^{*\rho\beta\top}$\quad$(\hat\theta_{\rm CLE}^{(t)}); \ddot\mL^{*\rho\beta}(\hat\theta_{\rm CLE}^{(t)}),\ddot\mL^{*\beta\beta}(\hat\theta_{\rm CLE}^{(t)})\Big)$.
Then, in the $(t+1)$-th iteration, the estimate of $\sigma_0^2$ can be obtained by solving \eqref{eq:CLEfsigma}, which is $\hat\sigma_{\rm CLE}^{2(t+1)}$. See Algorithm \ref{al:CLE} for details.
This leads to the final
estimator $\hat\theta_{\rm CLE}^{(t)}$. We establish the numerical convergence of $\hat\theta_{\rm CLE}^{(t)}$ in the next section.

\begin{algorithm}[h]
\caption{The iterative algorithm for the corrected likelihood estimator}
\begin{algorithmic}
\STATE \textbf{Input}: Initial estimator $ \hat\theta_{\rm CLE}^{(0)} =\big(\hat\gamma_{\rm CLE}^{(0)\top},(\hat\sigma^2_{\rm CLE})^{(0)}\big)^\top $, observed response $\mY^*$, observed covariates $\mX^*$, weighting matrix $W$, and privacy-protection level $\lambda^2$ and $\lambda_x^2$; \\
$t\leftarrow$0;\\
\REPEAT
\STATE \textbf{(Calculating Step)} Compute $\dot\mL^{*\gamma}(\hat\theta_{\rm CLE}^{(t)})$ and $\ddot\mL^{*\gamma}(\hat\theta_{\rm CLE}^{(t)})$ based on $\mX^*$ and $\mY^*$.
\STATE \textbf{(Debiasing Step)} Compute $\Delta \textbf{S}_{\rm CL}^{(t)}$ by \eqref{eq:CLEfrho}--\eqref{eq:CLEfsigma} and $\Delta \textbf{H}_{\rm CL}^{(t)}$ by \eqref{eq:CLEs}.\\
\STATE \textbf{(Updating Step)} Obtain $\hat\theta_{\rm CLE}^{(t+1)}=\Big(\hat\gamma_{\rm CLE}^{(t+1)\top},(\hat\sigma^2_{\rm CLE})^{(t+1)}\Big)^\top$ as,
\beqrs
\hat\gamma_{\rm CLE}^{(t+1)}&=&\hat\gamma_{\rm CLE}^{(t)}- \Big\{\ddot\mL^{*\gamma}(\hat\theta_{\rm CLE}^{(t)})- \Delta \textbf{H}_{\rm CL}^{(t)} \Big\}^{-1}\Big\{\dot\mL^{*\gamma}(\hat\theta_{\rm CLE}^{(t)}) - \Delta \textbf{S}_{\rm CL}^{(t)}  \Big\},\\
(\hat\sigma^2_{\rm CLE})^{(t+1)}&=&N^{-1}\Bigg[\Big\{S(\hat\rho_{\rm CLE}^{(t+1)})\mathbf{Y}^*-\mathbf{X}^*\hat\beta_{\rm CLE}^{(t+1)}\Big\}^\top\Big\{S(\hat\rho_{\rm CLE}^{(t+1)})\mathbf{Y}^*-\mathbf{X}^*\hat\beta_{\rm CLE}^{(t+1)}\Big\}\n\\
&&-\lambda^2\tr\Big\{S(\hat\rho_{\rm CLE}^{(t+1)})S(\hat\rho_{\rm CLE}^{(t+1)})^\top\Big\}\Bigg]-\lambda_x^2\hat\beta^{(t+1)\top}_{\rm CLE,2}\hat\beta^{(t+1)}_{\rm CLE,2},
\eeqrs
where $\hat\beta^{(t+1)}_{\rm CLE,2}$ is the CLE for $\beta_{02}$ in the $(t+1)$-th iteration.
\STATE $t \leftarrow t+1;$
\UNTIL $\|\hat\theta_{\rm CLE}^{(t+1)}-\hat\theta_{\rm CLE}^{(t)}\|<10^{-6}$.
\STATE \textbf{Output}: Corrected likelihood estimator $\hat\theta_{\rm CLE}^{(t)}$.
\end{algorithmic}\label{al:CLE}
\end{algorithm}

\subsection{Corrected least squares estimator}
The proposed CLE  offers a feasible parameter-estimation method for the PSAR model.
However, the calculation of the CLE involves the determinants of $S=I_N-\rho W$ and $\Omega=\sigma^2I_N+\lambda^2 SS^{\top}$. This makes the estimation process computationally expensive for large-scale networks with complexity of  order $O(N^3)$. Consequently, inspired by the neighborhood selection approach for high-dimensional graphs \citep{meinshausen2006variable,peng2009partial}, and the least squares estimation method by \cite{huang2019} and \cite{zhu2020multivariate}, we propose a corrected least squares estimator for the PSAR model as follows.

To illustrate the idea, we start with model \eqref{sarp} based on the true response $\mY$ and covariates $\mX_2$. Consider the conditional expectation of $Y_i$, given the responses of all other nodes. Define $\mF_{(-i)}=\sigma\{(X_{i'},Y_{i'}):i'\neq i\}$ to be the $\sigma$-field generated by all other nodes. Then, under the assumption of a normally distributed $\mE$, we have,
$E\{Y_{i}|\mF_{(-i)}\}=\mu_{i}+\sum_{j\neq i}^{N}\alpha_{ij}(Y_{j}-\mu_{j})$, where $\mu_{i}=E(Y_i)$ and
\beqr\label{eq:conexp}
\alpha_{ij}=\frac{\rho_0(\omega_{ij}+\omega_{ji})-\rho_0^2\sum_{k}\omega_{ki}\omega_{kj}}{1+\rho_0^2\sum_{k}\omega_{ki}^2}.
\eeqr
Define $d_\rho=\diag^{-1}(S^\top S)$ and recall that $\gamma=(\rho,\beta^\top)^\top$.
%
Accordingly, a least-squares-type objective function can be constructed as,
\beqr\label{eq:obj}
\mL_{\rm LS}(\gamma)=\mL_{\rm LS}(\gamma;\mY,\mX)=\sum_i\Bigg[Y_i-E\Big\{Y_i\Big|\mF_{(-i)}\Big\}\Bigg]^2=\Bigg\Vert d_\rho S^{\top}\Big(S\mY-\mX\beta\Big)\Bigg\Vert^2.
\eeqr
Then, a least squares estimator can be obtained as $\hat\gamma_{\rm LS}=\argmin \mL_{\rm LS} (\gamma)$.

We make two remarks about this method. First, from equation \eqref{eq:conexp}, for node $i$, only directly connected nodes (i.e., $a_{ij}+a_{ji}\neq 0$) and certain types of indirectly connected nodes with $\sum_{k}a_{ki}a_{kj}\neq 0$ are considered for the calculation of the conditional mean. These nodes will then further be involved in the computation of the objective function \eqref{eq:obj}. Thus, as long as the network is sufficiently sparse, the computation of the objective function will be efficient. Second, the construction of the objective function here is inspired by the assumption that $\mE$ follows a normal distribution. However, the formal statistical analysis does not rely on the normal assumption. Specific assumptions of the error term are provided in the next section.

However, the key challenge here is that we can only observe the noise-added responses $\mY^*$ and covariates $\mX_2^*$. This leads to the noise-added loss function $\mL_{\rm LS}^*(\gamma)=\mL_{\rm LS}(\gamma;\mY^*,\mX^*)$ instead of $\mL_{\rm LS}(\gamma)$. Similar to the analysis of the CLE, we consider a corrected least squares method, while simultaneously correcting the biases introduced by the observational errors of $\mX^*$ and $\mY^*$ on both the first- and second-order derivatives. This results in the {\it corrected least squares estimator} (CLS) $\hat\gamma_{\rm CLS}$, which is the solution to $\dot \mL_{\rm LS}^*(\gamma)-\Delta \textbf{S}_{\rm LS}(\gamma)=\mathbf{0}_{p+1}$.

Specifically, let $\dot \mL^*_{\rm LS}(\gamma) = \big(\dot \mL^{*\rho}_{\rm LS}(\gamma),\dot \mL^{*\beta}_{\rm LS}(\gamma)^\top\big)^\top\in\mR^{p+1}$ be the first-order derivative of $\mL_{\rm LS}(\gamma)$ with respect to $\gamma$. Here, $\dot \mL^{*\rho}_{\rm LS}(\gamma)$ and $\dot \mL^{*\beta}_{\rm LS}(\gamma)$ represent the first-order derivatives of $\mL_{\rm LS}(\gamma)$ with respect to $\rho$ and $\beta$, respectively.
Define $\dot d_\rho$ and $\ddot d_\rho$ to be the first- and second-order derivatives of $d_\rho$ with respect to $\rho$, respectively. Define $ \mW=W^{\top}S+S^{\top}W$ and $\mathbf{V}^*=S\mY^*-\mX^*\beta$. It can be verified that $E\big\{\dot\mL_{\rm LS}(\gamma_0)\big\}=\mathbf{0}_{p+1}$, and
\beqr
E\big\{\dot \mL^{*\rho}_{\rm LS}(\gamma_0)\big\}&=&E\Bigg[-2\Big(d_{\rho_0} S_0^\top \mV_0^*\Big)^\top\Big(d_{\rho_0} W^\top \mV_0^*+d_{\rho_0} S_0^\top W\mY^*-\dot d_{\rho_0} S_0^\top \mV_0^*\Big)\Bigg]\n\\ \nonumber
&=&2\lambda^2\Big[\tr\big\{(S_0^\top S_0)^2 d_{\rho_0} \dot d_{\rho_0} \big\}-\tr\big(S_0^\top S_0 d^2_{\rho_0}\mW_0\big)\Big]\\
&&+2\lambda_x^2\beta_{02}^\top\beta_{02}\Big\{\tr\big( S_0^\top S_0 d_{\rho_0} \dot d_{\rho_0}  \big)-\tr\big(S_0 d^2_{\rho_0}W^\top\big)\Big\}\label{eq:firder},\\
E\big\{\dot \mL^{*\beta}_{\rm LS}(\gamma_0)\big\}&=&E\Big(-2\mX^{*\top} S_0d_{\rho_0}^2S_0^\top \mV_0^*\Big)=\left(\mathbf{0}_{p_1}^\top,2\lambda_x^2\tr(S_0^\top S_0d_{\rho_0}^2)\beta_{02}^\top\right)^\top.\label{eq:firder0}
\eeqr
The verification details of equations \eqref{eq:firder} and \eqref{eq:firder0} are given in Appendix A.2. Define the right-hand side of \eqref{eq:firder}-\eqref{eq:firder0} as $ \Delta \textbf{S}_{\rm LS}(\gamma_0)=E\big\{\dot \mL_{\rm LS}^*(\gamma_0)\big\}$. Consequently, the estimator obtained by directly minimizing $\mL_{\rm LS}^*(\gamma)$ cannot be consistent.


To propose a Newton-Raphson-type algorithm, we still need to discuss the second-order derivatives. Define $\ddot\mL_{\rm LS}(\gamma)=\big(\ddot\mL_{\rm LS}^{\rho\rho}(\gamma),\ddot\mL_{\rm LS}^{\rho\beta}(\gamma)^\top;\ddot\mL_{\rm LS}^{\rho\beta}(\gamma),\ddot\mL_{\rm LS}^{\beta\beta}(\gamma)\big)$ to be the second-order derivative of $\mL_{\rm LS}(\gamma)$.
We can further define the difference between the expectations of second-order derivatives for $\mL^*_{\rm LS}(\gamma)$ and $\mL_{\rm LS}(\gamma)$ as,
\beqr\label{eq:CLSs}
\Delta\mathbf{H}_{\rm LS}(\gamma_0)=E\big\{\ddot\mL^*_{\rm LS}(\gamma_0)\big\}-E\big\{\ddot\mL_{\rm LS}(\gamma_0)\big\}=\left(
           \begin{array}{cc}
             \mathbf{H}_{\rm LS}^{\rho\rho} & \mathbf{H}_{\rm LS}^{\rho\beta\top} \\
             \mathbf{H}_{\rm LS}^{\rho\beta} & \mathbf{H}_{\rm LS}^{\beta\beta}
           \end{array}
         \right).
\eeqr
\noindent
As a result, we can verify that $\Delta\mathbf{H}_{\rm LS}(\gamma_0)$ is symmetric and
\beqr
\mathbf{H}_{\rm LS}^{\rho\rho}&=&2\lambda^2\Bigg[\tr\Big(\mW_0 d_{\rho_0}^2 \mW_0\Big)+2\tr\Big(S_0^\top S_0 d_{\rho_0}^2 W^\top W\Big)-4\tr\Big(\mW_0 d_{\rho_0} \dot d_{\rho_0} S_0^\top S_0\Big)\n\\
&&+\tr\Big\{\Big(S_0^\top S_0\Big)^2 \Big(\dot {d}^2_{\rho_0}+d_{\rho_0} \ddot d_{\rho_0}\Big) \Big\}\Bigg]+2\lambda_x^2\beta_{20}^\top\beta_{20}\Big[\tr\big\{W^\top Wd_{\rho_0}^2-4\tr(Wd_{\rho_0}\dot{d_{\rho_0}}S_0^\top)\label{eq:seccr}\\
&&+S_0^\top S_0(\ddot{d_{\rho_0}}d_{\rho_0}+\dot{d}^2_{\rho_0})\big\}\Big],\n\\
\mathbf{H}_{\rm LS}^{\rho\beta}&=&\Big({\bf0}_{p_1}, 4\lambda_x^2\tr(S_0d_{\rho_0}\dot d_{\rho_0}S_0^\top)\beta_{02}^\top-4\lambda_x^2\tr(Wd^2_{\rho_0}S_0^\top)\beta_{02}^\top\Big)^\top,\n\\
\mathbf{H}_{\rm LS}^{\beta\beta}&=&\begin{pmatrix} {\bf 0}_{p_1\times p_1}& {\bf 0}_{p_1\times p_2}\\
{\bf 0}_{p_1\times p_2}^\top& 2\lambda_x^2\tr(S_0 d_{\rho_0}^2 S_0^\top)I_{p_2} \end{pmatrix}.\n
\eeqr
\noindent
The verification details of \eqref{eq:seccr} and the expressions for it are given in Appendix A.2.

An iterative algorithm (i.e., Algorithm \ref{al:CLS}) can then be developed, which can also be described in the three steps (calculating, debiasing, and updating). Here, we only list the updating equation for simplicity.  Let $\hat\gamma_{\rm CLS}^{(0)}=(\hat\rho_{\rm CLS}^{(0)},\hat\beta_{\rm CLS}^{(0)\top})^\top\in\mR^{p+1}$ be the initial estimator and $\hat\gamma_{\rm CLS}^{(t)}=(\hat\rho_{\rm CLS}^{(t)},\hat\beta_{\rm CLS}^{(t)\top})^\top\in\mR^{p+1}$ be the estimator obtained in the $t$-th
iteration. The estimator in the ($t+1$)-th iteration can then be updated as follows,
\beqr
\hat\gamma_{\rm CLS}^{(t+1)}=\hat\gamma_{\rm CLS}^{(t)}-\Big\{\ddot \mL_{\rm LS}^*\Big(\hat\gamma_{\rm CLS}^{(t)}\Big)-\Delta\textbf{H}_{\rm LS}^{(t)}\Big\}^{-1}\Big\{\dot \mL_{\rm LS}^*\Big(\hat\gamma_{\rm CLS}^{(t)}\Big)-\Delta\textbf{S}_{\rm LS}^{(t)}\Big\},\label{eq:CLS_iter}
\eeqr
where $\Delta\textbf{S}_{\rm LS}^{(t)}=\Delta\textbf{S}_{\rm LS}\Big(\hat\gamma^{(t)}_{\rm CLS}\Big)$ and $\Delta\textbf{H}_{\rm LS}^{(t)}=\Delta\textbf{H}_{\rm LS}\Big(\hat\gamma^{(t)}_{\rm CLS}\Big)$. By the time of convergence, we obtain the final estimator. It will be shown in the next section that $\hat\gamma_{\rm CLS}^{(t)}$ numerically converges to $\hat\gamma_{\rm CLS}$. This is the second estimator we have developed in this work.

\begin{algorithm}[h]
\caption{The iterative algorithm for the corrected least squares estimator}
\begin{algorithmic}
\STATE \textbf{Input}: Initial estimator $\hat\gamma_{\rm CLS}^{(0)}$, observed response $\mY^*$, observed covariates $\mX^*$, weighting matrix $W$, privacy-protection level $\lambda^2$ and $\lambda_x^2$; \\
$t\leftarrow$0;\\
\REPEAT
\STATE \textbf{(Calculating Step)} Compute $\dot\mL^*_{\rm LS}(\hat\gamma_{\rm CLS}^{(t)})$ and $\ddot\mL^*_{\rm LS}(\hat\gamma_{\rm CLS}^{(t)})$ based on $\mX^*$ and $\mY^*$.
\STATE \textbf{(Debiasing Step)} Compute $\Delta \textbf{S}_{\rm LS}^{(t)}$ by \eqref{eq:firder}--\eqref{eq:firder0} and $\Delta \textbf{H}_{\rm LS}^{(t)}$ by \eqref{eq:CLSs}.\\
\STATE \textbf{(Updating Step)}  Obtain $\hat\gamma_{\rm CLS}^{(t+1)}$ by \eqref{eq:CLS_iter}.
\STATE $t \leftarrow t+1;$
\UNTIL $\|\hat\gamma_{\rm CLS}^{(t+1)}-\hat\gamma_{\rm CLS}^{(t)}\|<10^{-6}$.
\STATE \textbf{Output}: Corrected least squares estimator $\hat\gamma_{\rm CLS}^{(t)}$.
\end{algorithmic}\label{al:CLS}
\end{algorithm}

%
In terms of the computational advantage, the CLS is observed to avoid the need for large-scale matrix-inversion calculations throughout the entire algorithm. This includes computing the first-order derivative, second-order derivative, and correction terms in equations \eqref{eq:firder}--\eqref{eq:seccr}. Consequently, this approach significantly reduces the computational time required compared to the CLE. The substantial computational advantages of $\hat\gamma_{\rm CLS}$ are further illustrated in the  numerical analysis.

\section{Theoretical properties} In this section, we first theoretically analyze both the numerical and statistical properties of the CLE and CLS. Then, the generalizations of PSAR are carefully presented to extend the method to more model forms, enabling support for a wider range of data analysis applications.
First, we introduce some theoretical assumptions.

\subsection{Theoretical assumptions} 
Define $\|B\|_1=\max_j\sum_i|b_{ij}|$ to be $L_1$ norm and $\|B\|_\infty=\max_i\sum_j|b_{ij}|$ to be $L_\infty$ norm for an arbitrary matrix $B=(b_{ij})$.
For theoretical analysis, the following regularity conditions are required.

(C1) (Covariates) Assume that $\lim_{N\to\infty} N^{-1} \mX^\top \mX$ exists and is non-singular.

(C2) (Uniform Boundedness) The elements of $\mX$ are uniformly bounded for all $N$. Moreover, the elements $w_{ij}$ of $W$ are uniformly bounded with the uniform order $O(1/h_n)$, where $h_n$ can be bounded or divergent with $h_n/N \to 0$ as $N \to \infty$. Furthermore, $W$ and $S_0^{-1}$ have bounded $L_1$ and $L_\infty$ norms.

(C3) (Noise Term) Assume that all the ${e_i}$s, $\vare_i$s, and $\vare _{ij}^x$s ($1\leq i,j\leq N$) are independently and identically distributed with mean zero. For the variance, we assume that $\text{Var}(\vare_i)=\lambda^2$ and $\text{Var}(\vare_{ij}^x)=\lambda_x^2$. To simplify the asymptotic covariance form of the estimators, we assume $E(e_i^3)=E(\vare_i^3)=E(\vare_{ij}^{x3})=0$. For higher moment conditions, we assume that $E(e_i^4)=\mu_4^e$, $E(\vare_i^4)=\mu_4^\vare$, $E(\vare_{ij}^{x4})=\mu_4^{\vare_x}$, $E(e_i^2-\sigma_0^2)^4=c'_e$, $E(\vare_i^2-\lambda^2)^4=c'_\vare$, and $E(\vare_{ij}^{x2}-\lambda_x^2)^4=c'_{\vare_x}$, for positive constants $\mu_4^e$, $\mu_4^\vare$, $\mu_4^{\vare_x}$, $c'_e,$ $c'_\vare$, and $c'_{\vare_x}$.

\noindent
Condition (C1) requires the covariance to exist and be non-singular. The same condition was adopted by \cite{Lee:2004}, \cite{yang2016}, and \cite{zhu2020multivariate}. Condition (C2) requires the uniform boundedness of features $\mX$, the weighting matrix $W$, and $S_0^{-1}$; this is a classical regularity condition in the SAR model \citep{Lee:2004,yang2016}. Condition (C3) is a typical assumption for the noise term \citep{huang2019, zhu2020multivariate, huang2020two}. It is remarkable that the artificially created noise is allowed to be non-Gaussian, as long as the moment conditions are satisfied. Furthermore, condition $E(e_i^3)=E(\vare_i^3)=E(\vare_{ij}^{x3})=0$ is assumed for convenience. The aim is to simplify the form of the asymptotic covariances for the estimators. Notably, the theory to be presented can be softly generalized under the assumption of non-zero third-order moments using the same theoretical proof techniques in this work.
Given these conditions, we then establish the theoretical results in the subsequent subsections.

\subsection{Theoretical properties of the CLE}
Before establishing the theoretical properties of the CLE, we first address the challenges in the theoretical analysis of the PSAR model compared to the traditional SAR models without privacy protection. Next, we discuss the identification issue of $\hat\theta_{\text{CLE}}$. The consistency and asymptotic normality of $\hat\theta_{\text{CLE}}$ are then carefully established. Finally, the numerical convergence of the proposed iterative algorithm is rigorously proved.

In the theoretical analysis of the traditional SAR estimators based on the likelihood function, the proof core involves handling of the matrix $S^{-1}=(I_N-\rho W)^{-1}$. As pointed out by \citep{Lee:2004}, the boundness assumption is fundamentally important for studying the asymptotic properties of estimators for spatial econometric models. Take classical SAR as an example, where $\mathbf{Y}=(I_N-\rho W)^{-1}\mathcal{E}$, the boundness of $L_1$- and $L_\infty$-norms guarantees that the variances of  each $Y_i$  with $1\leq i\leq N$ is bounded as $N$ goes to infinity. The existing literature presents two primary approaches to address this issue. The first method considers bounding both the $L_1$ norm and $L_\infty$ norm of $S^{-1}$. This can be verified by assuming that the row and column sums of $W$ and $S^{-1}$ are uniformly bounded; for example, refer to \cite{Lee:2004} and \cite{yang2016}.
Meanwhile, the second approach considers the application of Taylor's expansion to $S^{-1}$, based on the assumption that the weighting matrix $W$ is a transition-probability matrix, and the stationary distribution for the network nodes exists; for example, refer to \cite{huang2019} and \cite{zhu2020multivariate}.

However, to estimate the parameters for the PSAR model based on the likelihood function, we need to analyze $\Omega^{-1}=\big(\sigma^2I_N+\lambda^2 SS^{\top}\big)^{-1}$, which distinguishes the estimation procedure from the previous two approaches. On the one hand, even if $W$ and $S^{-1}$ satisfy the boundedness conditions of the $L_1$ and $L_\infty$ norms, it cannot be directly proven that $\Omega^{-1}$ is bounded in  $L_1$ and $L_\infty$ norms. Consequently, the first method cannot be employed. On the other hand, notice that $S^{\top}S$ in $\Omega^{-1}$ cannot be analyzed as a transition-probability matrix; thus, the assumption of the existence of a stationary distribution cannot be made either. This makes the second method inapplicable.
Therefore, we need to reexamine the theoretical properties of the CLE $\hat\theta_{\text{CLE}}$ based on the PSAR model. We begin with the discussion of the identification issue. Before presenting the theoretical results, the following identification condition is required.

(C4) (Identification) Assume that at least one of the following conditions holds: (a) $\lim_{N \to \infty} N^{-1} \beta_0^\top \mX^\top (WS_0^{-1})^\top \Omega^{-1} H_{\Omega} WS_0^{-1} \mX \beta_0$ is positive for any $(\rho,\sigma^2)$ in parameter space; or (b) $\lim_{N \to \infty} N^{-1}  \beta_0^\top \mX^\top (WS_0^{-1})^\top \Omega^{-1}H_{\Omega} WS_0^{-1} \mX \beta_0 = 0$, sequence $\{h_n\}$ is bounded, and for any $(\rho,\sigma^2) \neq (\rho_0,\sigma_0^2)$,
${
\lim_{N\rightarrow \infty}}N^{-1} \tr \big\{S^\top \Omega^{-1} S S_0^{-1} \Omega_0 (S_0^{-1})^\top \big\} \neq 1.
$

\noindent
Condition (C4) assumes that either (a) covariates $\mX$ and $WS_0^{-1}\mX\beta_0^\top$ do not exhibit asymptotic multicollinearity, or (b) the variance matrix of $\mY^*$ is unique. A similar type of condition was also assumed by \cite{Lee:2004}.
Then, we have the following theorem.
\begin{theorem}[Identification of CLE]
\label{thm:CLE_identification} 
Assume (C1)--(C4) hold. Then $ \theta_0$ is identifiable.
\end{theorem}
\noindent
The proof of Theorem \ref{thm:CLE_identification} is given in Appendix C.1.
We then examine the statistical property of the CLE, which results in the following theorem.
\noindent
\begin{theorem}[Asymptotic Normality of CLE]
\label{thm:cle}
Assume that (C1)--(C4) hold. Then, we have
$
\sqrt{N}(\hat \theta_{\rm CLE} - \theta_0) \to_d N(\mathbf{0}_{p+2}, (\Sigma_2^{\rm CL})^{-1} \Sigma_1^{\rm CL} (\Sigma_2^{\rm CL})^{-1})
$
as $N \to \infty$, where $\Sigma_1^{\rm CL}$ and $\Sigma_2^{\rm CL}$ are assumed to be positive definite matrices expressed as
\beqr\label{eq:CLE_Cov}
\Sigma_2^{\rm CL}=\begin{pmatrix}
\Sigma^{\rm CL}_{\rho\rho}&(\Sigma^{\rm CL}_{\rho\beta})^\top&\Sigma^{\rm CL}_{\rho\sigma^2}\\
\Sigma^{\rm CL}_{\rho\beta}&\Sigma^{\rm CL}_{\beta\beta}&\textbf{0}_{p\times1}\\
\Sigma^{\rm CL}_{\rho\sigma^2}&\textbf{0}_{1\times p}&\Sigma^{\rm CL}_{\sigma^2\sigma^2}\\
\end{pmatrix}
,\Sigma_1^{\rm CL}=\Sigma_2^{\rm CL}+\begin{pmatrix}
\Delta^{\rm CL}_{\rho\rho}&(\Delta^{\rm CL}_{\rho\beta})^\top&\Delta^{\rm CL}_{\rho\sigma^2}\\
\Delta^{\rm CL}_{\rho\beta}&\Delta^{\rm CL}_{\beta\beta^\top}&\Delta^{\rm CL}_{\beta\sigma^2}\\
\Delta^{\rm CL}_{\rho\sigma^2}&(\Delta^{\rm CL}_{\beta\sigma^2})^\top&\Delta^{\rm CL}_{\sigma^2\sigma^2}\\
\end{pmatrix}.
\eeqr
The formula of the asymptotic covariance in \eqref{eq:CLE_Cov} is provided in Appendix A.2.
\end{theorem}
\noindent
The proof of Theorem \ref{thm:cle} is given in Appendix C.2. The theorem guarantees the asymptotic normality of $\hat \theta_{\text{CLE}}$. By combining the previous discussion on identification issues,
we can employ the proposed Newton-Raphson-type method to iteratively obtain the CLE by Algorithm \ref{al:CLE}. In real practice, we obtain $\hat\theta_{\text{CLE}}^{(t)}$ instead of $\hat \theta_{\text{CLE}}$. Thus, it is necessary to discuss the relationship between $\hat \theta_{\text{CLE}}$ and $\hat\theta_{\text{CLE}}^{(t)}$ in the proposed algorithm. In this regard, let $\lambda_{\min}(B)$ and $\lambda_{\max}(B)$ denote the smallest and largest eigenvalues of a symmetric matrix $B$, respectively. Then the following theorem can be established.

\begin{theorem}[Numerical Convergence of CLE]
\label{thm:LCE_numerical}
Assume that (C1)--(C4) hold. Further assume that the initial value $\hat \theta_{\CLE}^{(0)}$ lies sufficiently close to $\theta_0$ in the sense that $\|\hat\theta_{\CLE}^{(0)} - \theta_0\|_{\delta} \leq (4 \lambda_{\max}(\Sigma_2^{\rm CL}) L_{\max})^{-1} \lambda_{\min}(\Sigma_2^{\rm CL})$, for some natural norm $\|\cdot\|_{\delta}$ and a smoothness constant $L_{\max}$; see Appendix C.3 for details. Then $\hat \theta_{\CLE}^{(t)} \to \hat \theta_{\CLE}$ as $t \to \infty$ with probability tending to $1$.
\end{theorem}
\noindent
The proof of Theorem \ref{thm:LCE_numerical} is given in Appendix C.3.
Theorem \ref{thm:LCE_numerical} guarantees the numerical convergence of the proposed algorithm. This suggests that the proposed iterative algorithm can obtain $\hat \theta_{\CLE}$ with asymptotic normality.

\begin{remark} (Bias-Corrected Estimator for Asymptotic Covariance)
When $\mX$ is unobserved, the asymptotic covariance in \eqref{eq:CLE_Cov} cannot be obtained directly. Define $\Sigma_k^{\text{CL}} = \Sigma_k^{\text{CL}}(\mX, \theta)$ $(k = 1,2)$, which is a function of both $\mX$ and and true parameter $\theta$. Users could only obtain $\mX^*$, which leads to a difference as $\Delta_k(\theta) = E\{\Sigma_k^{\text{CL}}(\mX^*, \theta)\} - \Sigma_k^{\text{CL}}(\mX, \theta)$. A bias-corrected estimator can thus be constructed as $\widehat{\Sigma}_k^{\text{CL}} = \Sigma_k^{\text{CL}}(\mX^*, \hat \theta_{\text{CLE}}) - \Delta_k(\hat \theta_{\text{CLE}})$. The consistency of $\widehat{\Sigma}_k^{\text{CL}}$ can be established using similar techniques to Lemma B.6 in the Appendix. The estimated asymptotic covariance matrix is given by $(\widehat{\Sigma}_2^{\text{CL}})^{-1}\widehat{\Sigma}_1^{\text{CL}}(\widehat{\Sigma}_2^{\text{CL}})^{-1}$, which could be adopted to compute the standard error of $\hat \theta_{\text{CLE}}$.
\end{remark}
\begin{remark} (Trade-off Between Privacy Protection and Statistical Efficiency)
To gain intuition into the privacy-protection level and statistical efficiency, consider a pure SAR case with $\beta = 0$, known $\sigma^2$, and Gaussian errors. Estimating the network effect $\rho$, a Taylor expansion on $\Sigma_k^{\text{CL}}$ ($k = 1, 2$) yields leading terms for approximating the asymptotic covariance. It follows that $\Delta_{\rho\rho}^{\rm CL} = o(1)$, and
\beqrs\Sigma_2^{\rm CL} (\Sigma_1^{\rm CL})^{-1} \Sigma_2^{\rm CL} = \frac{\sigma^4}{N(\lambda^2 + \sigma^2)^2} \tr(W^2 + WW^\top) + o(1).
\eeqrs
We know that as privacy budget $\varepsilon$ decreases, the noise level $\lambda^2$ increases, 
As $\lambda^2$ (the privacy-protection level) increases, 
thus the asymptotic variance increases, indicating reduced statistical efficiency. 
Thus, a higher privacy-protection level enhances data security but decreases estimation eﬀiciency. In practice, data providers select a reasonable privacy-protection level to ensure data security while maintaining statistical efficiency. For more general cases, the explicit asymptotic variance becomes difficult to derive as $\lambda^2$ varies, and further evidence is provided through numerical simulations. 
\end{remark}
\subsection{Theoretical properties of the CLS}
In this subsection, we establish the theoretical properties of the CLS.
The analysis procedure is similar to those in Section 3.2. Therefore, for brevity and to avoid repetition, we will omit the detailed discussion of the technique here. We start with the identification issue. To this end, define $\mX_s=WS_0^{-1}\mX\beta_0$ and $\widetilde \mX=(\mX_s,\mX)\in\mR^{N\times(p+1)}$. The following assumption is necessary for the identification of the parameter.

(C4$^*$) (Identification) Assume that (a) $\lim_{N\to\infty}N^{-1} \widetilde \mX^\top  \widetilde \mX$ exists and is non-singular, and $\delta>0$ exists such that $\min_{|\rho|\leq 1-\delta}\lambda_{\min}(S S^\top)\geq \tau$, where $\tau$ is a positive constant; or (b) for the SAR model with no exogenous covariates, assume $I_N$, $W$, $W^\top$, and $W^\top W$ are linearly independent.

\noindent
Here, the identification issue is discussed in two cases: with covariates and without covariates. The condition is similar to that in \cite{zhu2020multivariate} and \cite{huang2020two}. We then have the following theorem.

\begin{theorem}[Identification of CLS]
\label{thm:CLS_identification}  
Assume (C1)--(C3) and (C4$^*$) hold.  Then, $\gamma_0$ is identifiable.
\end{theorem}
\noindent
The proof of Theorem \ref{thm:CLS_identification} is given in Appendix C.4. Next, we establish the following numerical and statistical theoretical properties of $\hat \gamma_{\text{CLS}}$.
\begin{theorem}[Numerical Convergence and Asymptotic Normality of CLS]
\label{thm:CLS}
 Assume that (C1)--(C3)  and (C4$^*$) hold. Further assume that the initial value $\hat \gamma_{\CLS}^{(0)}$ lies sufficiently close to $\gamma_0$ in the sense that $\|\hat \gamma_{\CLS}^{(0)} - \gamma_0\|_2 \leq 1/(2\lambda_{\max}(\Sigma_2^{\rm LS}) L_{\max}^{\prime})$, for some smoothness constant $L_{\max}^{\prime}$; see Appendix C.5 for details. Then $\hat\gamma_{\CLS}^{(t)} \to \hat \gamma_{\CLS}$ as $t \to \infty$ with probability tending to 1. And we further have
$\sqrt{N}(\hat \gamma_{\CLS} - \gamma_0) \to_d N(\mathbf{0}_{p+1}, (\Sigma_2^{\LS})^{-1} \Sigma_1^{\LS} (\Sigma_2^{\LS})^{-1})
$
as $N \to \infty$, where $\Sigma_1^{\LS}$ and $\Sigma_2^{\LS}$ are assumed to be positive definite matrices expressed as
\beqr\label{eq:CLS_Cov}
\Sigma_1^{LS}=\begin{pmatrix}
\Sigma^{LS}_{1\rho\rho}&(\Sigma^{LS}_{1\rho\beta})^\top\\
\Sigma^{LS}_{1\rho\beta}&\Sigma^{LS}_{1\beta\beta}\\
\end{pmatrix},
\Sigma_2^{LS}=\begin{pmatrix}
\Sigma^{LS}_{2\rho\rho}&(\Sigma^{LS}_{2\rho\beta})^\top\\
\Sigma^{LS}_{2\rho\beta}&\Sigma^{LS}_{2\beta\beta}\\
\end{pmatrix}.
\eeqr
See Appendix A.2 for the detailed formula of the asymptotic covariance in \eqref{eq:CLS_Cov}.
\end{theorem}
\noindent
Through Theorem \ref{thm:CLS}, we can conclude that the CLS estimator $\hat \gamma_{\text{CLS}}$ is $\sqrt{N}$-consistent, which has the same convergence rate as the CLE.

For statistical inference, similar to the discussion for $\hat \theta_{\text{CLE}}$, we can also provide consistent estimators for each element in the asymptotic covariance matrix \eqref{eq:CLS_Cov} based on the observed $\mX^*$.  Rewrite  $\Sigma_k^{\text{LS}}=\Sigma_k^{\text{LS}}(\mX,\gamma)$ $(k=1,2)$ as functions of $\mX$ and $\gamma$, which are expressed in \eqref{eq:CLS_Cov}. Then, we could plug $\mX^*,\hat \gamma_{\text{CLS}}$ in to obtain $\Sigma_k^{\text{LS}}(\mX^*,\hat \gamma_{\text{CLS}})$ for $k=1,2$. As a result, $\Delta_k^{\text{LS}}(\gamma)=E\{\Sigma_k^{\text{LS}}(\mX^*,\theta)\}-\Sigma_k^{\text{LS}}(\mX,\theta)$ could be calculated. See Appendix A.2 for a detailed expression of $\Delta_k^{\text{LS}}(\gamma)$.
With $\hat \gamma_{\text{CLS}}$ plugged in, the consistent estimators for $\Sigma_k^{\text{LS}}$ could be obtained as $\widehat{\Sigma}_k^{\text{LS}}=\Sigma_k^{\text{LS}}(\mX^*,\hat \gamma_{\text{CLS}})-\Delta_k^{\text{LS}}(\hat \gamma_{\text{CLS}})$ $(k=1,2)$. In this way, we could obtain the estimated standard error of $\hat \gamma_{\text{CLS}}$ by the estimated asymptotic covariance matrix $(\widehat{\Sigma}_2^{\text{LS}})^{-1}\widehat{\Sigma}_1^{\text{LS}}(\widehat{\Sigma}_2^{\text{LS}})^{-1}$. Using a technique similar to that used in the proof of Lemma B.8 in Appendix, the consistency of $\widehat{\Sigma}_k^{\text{LS}}$ can be established.

\subsection{Model extensions and related theoretical discussions}

The previously discussed model \eqref{sar_dp_ge} represents a simple form of the SAR model. In real applications, various extensions can be considered. As examples in this work, we explore: (1) extending the privacy-protected model to a multi-mode multivariate SAR model, which supports multivariate responses \citep{yang2016,zhu2020multivariate} and multiple types of nodes \citep{huang2020two}; (2) introducing perturbations to the network structure to assess their impact on the robustness and reliability of privacy-preserving techniques \citep{Lewbel2024}.

\textbf{Privacy-Protected Multi-Mode Multivariate SAR}.
We consider extending the PSAR model to the privacy-protected multi-mode multivariate SAR (MMSAR) model, which can be adopted to describe the interaction of responses between different types of nodes. Consider a multi-mode network with $N_k$ nodes in the $k$th group ($1\leq k\leq K$), and $N=\sum_{k}N_k$.
The adjacency matrix between nodes of type $k_1$ and type $k_2$ is $A_{k_1k_2}=(a_{k_1k_2,i_1i_2})\in\mathbb{R}^{N_{k_1}\times N_{k_2}}$, where $a_{k_1k_2,i_1i_2} = 1$ if there is a edge from $i_1$ to $i_2$ and $a_{k_1k_2,i_1i_2} = 0$ otherwise $(1\leq i_1\leq N_{k_1},1\leq i_2\leq N_{k_2}, 1\leq k_1,k_2\leq K)$. Then define $W_{k_1k_2}$ is a row-normalized weighting matrix of $A_{k_1k_2}(1\leq k_1,k_2\leq K)$.
Suppose there are multivariate responses $\mY_k = (Y_{k,ij})\in \mR^{N_k\times q_k}$ $(1\leq i\leq N_k, 1\leq j\leq q_k)$ for the nodes in the $k$th group and error matrix $\widetilde{\mE}_k=(e_{k,ij}) \in \mR^{N_k\times q_k}$. Define $\mX_k=(\mathbf{x}_{k,1}^\top, \cdots,\mathbf{x}_{k,N_k}^\top)^\top\in\mR^{ N_k\times p_k}$ as the exogenous covariates for the $k$th group.
Then the multi-mode multivariate spatial autoregressive model can be defined as
\beqr{\label{eq:MMSAR}}
\mY_{k}=\sum_{k_1=1}^{K} W_{kk_1}\mY_{k_1}D_{k_1k}+\mX_{k}B_{k}+\widetilde{\mE}_{k}\quad(k=1,...,K),
\eeqr
where $D_{k_1k}=(d_{k_1k,j_1j})\in\mathbb{R}^{q_{k_1}\times q_{k}}$ is the network effect parameter matrix, with $d_{k_1k,j_1j}$ representing the impact of the $j_1$-th $(1\leq j_1\leq q_1)$ response of the $k_1$-th node type on the $j$-th $(1\leq j\leq q)$ response of the $k$-th node type. This is referred to as the network \textit{cross-mode effect} when $k_1 \neq k$ and \textit{within-mode effect} when $k_1 = k$. The matrix $B_{k}\in\mathbb{R}^{p_{k}\times q_{k}}$ corresponds to the covariate parameters for $\mX_{k}$.

Consider the vector form of model \eqref{eq:MMSAR}. Define $Q=\sum_k N_k q_k$ and $P=\sum_k p_k q_k$. Let
$\mathcal{Y} = \{{\rm vec}^\top(\mY_1),...,{\rm vec}^\top(\mY_K)\}^\top\in\mR^{Q}$, $\mathcal{X} = \diag(I_{q_1}\otimes\mX_1,\dots, I_{q_K}\otimes\mX_K)\in\mathbb{R}^{Q\times P}$, $\beta=\{{\rm vec}^\top(B_1),\dots,{\rm vec}^\top(B_K)\}^\top\in\mR^{P}$, and $\mE=\{{\rm vec}^\top(\widetilde{\mE}_1),\dots,{\rm vec}^\top(\widetilde{\mE}_K)\}^\top\in\mR^{Q}$.
Then, the model can be written as
$\mathcal{Y}=\mW_D \mathcal{Y}+\mathcal{X}\beta+\mE$ with each component of $\mW_D\in\mR^{Q\times Q}$ denoted as $D_{k_1k_2}^\top\otimes W_{k_2k_1}$ for $1\leq k_1,k_2\leq K$.
In privacy-protected scenarios, the observed data is given by $\mathcal{Y}^*= \mathcal{Y} + \mathcal{E}$, $\mathcal{X}^*=\mathcal{X}+\mathcal{E}_x$, where $\mathcal{E}_x = \diag(I_{q_1}\otimes\mathcal{E}_{x1},\dots,I_{q_K}\otimes\mathcal{E}_{xK})\in\mathbb{R}^{Q\times P}$, with $\mathcal{E}$ and $\mathcal{E}_{xk}$ $(1\leq k\leq K)$ added independently with zero mean and variance matrices $\lambda^2I_{Q}$ and $\lambda_x^2I_{P}$, respectively. Define $\gamma_{MM} \in \mathbb{R}^{\widetilde{Q}}$ with $\widetilde{Q}=\sum_{k_1, k_2}q_{k_1}q_{k_2}+\sum_k p_k q_k$ as the parameter vector that incorporates all the information from \(D_{k_1,k_2}\) and \(B_k\).
We construct the objective function and obtain the CLE or CLS estimator for $\gamma_{MM}$ by correcting for first- and second-order derivatives to deal with observational errors in $\mathcal{X}^*$ and $\mathcal{Y}^*$. Consequently, we can establish the following corollary.

\begin{corollary}
\label{thm:MMSAR_CLE}
(Asymptotic Normality for Privacy-Protected MMSAR Estimation) Assume that (C1)--(C3) hold. Denote $\Sigma_1^{MM,\CLS}$, $\Sigma_2^{MM,\CLS}$, $\Sigma_1^{MM,\CLE}$, and $\Sigma_2^{MM,\CLE}$ as positive definite matrices, with definitions given in Appendix D. As $N\rightarrow \infty$:
(1) If (C4) holds:
$\sqrt{N}(\hat \gamma_{MM,\CLE} - \gamma_{MM}) \to_d N(\mathbf{0}_{\widetilde{Q}}, (\Sigma_2^{MM,\CLE})^{-1} \Sigma_1^{MM,\CLE} (\Sigma_2^{MM,\CLE})^{-1});$
(2) If (C4$^*$) holds:
$\sqrt{N}(\hat \gamma_{MM,\CLS} - \gamma_{MM}) \to_d N(\mathbf{0}_{\widetilde{Q}}, (\Sigma_2^{MM,\CLS})^{-1} \Sigma_1^{MM,\CLS} (\Sigma_2^{MM,\CLS})^{-1}).$
\end{corollary}

\textbf{Privacy-Protected SAR with Perturbed Network}. As a further discussion of the PSAR model, we consider randomly perturbing the edges in the network to protect the privacy of the network structure. The true network is represented by \(A\), and we define the noise-perturbed adjacency matrix as \(A^*\). In this way, we obtain the observed weighting matrix \(W^*\), which is normalized from \(A^*\). Consequently, we calculate the corrected least squares estimator \(\hat\gamma^*_{\rm CLS}\) based on \(W^*\) instead of \(\hat\gamma_{\rm CLS}\) based on \(W\). However, we cannot allow a large number of edges in the network to change, as excessive noise would prevent us from obtaining consistent parameter estimates. To address this issue, and inspired by \cite{Lewbel2024}, we propose the following constraint condition (C5). Based on (C5), we establish the theoretical properties of the CLS under the condition of perturbed network edges. This corollary indicates that once condition (C5) is satisfied, we can still ensure the consistency and asymptotic normality of the CLS estimator.

(C5) (Perturbed Network) Assume \(\sum_i\sum_jE(|A^*_{ij}-A_{ij}|)=O(N^s)\) for some positive constant \(s < 1/2\), and assume that \(W^*\) is uniformly bounded in both row and column sums in probability.

\begin{corollary}
\label{thm:MissW_CLS}
(Asymptotic Normality of CLS For Perturbed Network)
Assume conditions (C1)--(C3), (C4$^*$), and (C5) hold. As $N \to \infty$, we then have
$\sqrt{N}(\hat \gamma^*_{\CLS} - \gamma_0) \to_d N(\mathbf{0}_{p+1},$\quad$ (\Sigma_2^{\LS})^{-1} \Sigma_1^{\LS} (\Sigma_2^{\LS})^{-1}),
$
where $\Sigma_1^{\LS}$ and $\Sigma_2^{\LS}$ are positive definite matrices defined in Theorem \ref{thm:CLS}.
See Appendix D for the detailed formula.
\end{corollary}

\section{Simulation studies}

\subsection{Simulation models}

To demonstrate the finite-sample performance of the proposed methods, we present three simulation examples based on the following generating mechanisms: the network structure $A$, data generation, noise distribution, and privacy-protection level. The classical QMLE (treating the observed variables as the true ones) and the proposed estimators (i.e., CLE and CLS) are compared. The network structures are given as follows.

\textbf{Example 1.} (Dyad Independence Network) Following \cite{holland1981exponential}, we define a dyad as $\mathbf{A}_{ij} = (a_{ij}, a_{ji})$ ($1\le i < j\le N$) and assume that different $\mathbf{A}_{ij}$s are independent. To allow for network sparsity, we set $P(\mathbf{A}_{ij} = (1,1)) = 10N^{-1}$ and $P(\mathbf{A}_{ij} = (1,0)) = P(\mathbf{A}_{ij} = (0,1)) = 0.5N^{-0.8}$. Then, the probability of a null dyad is $P(\mathbf{A}_{ij} = (0,0)) = 1-10N^{-1}-N^{-0.8}$, which is close to 1 when $N$ is large.

\textbf{Example 2.} (Stochastic Block Network) The next network type that is considered is the stochastic block network \citep{wang1987stochastic,nowicki2001estimation}. Following \cite{nowicki2001estimation}, we randomly assign a block label $k$ ($1\leq k\leq K$) for each node with $K= 20$ as the total number of blocks. Define $P(a_{ij}=1) = 20N^{-1}$ if $i$ and $j$ belong to the same block, and $P(a_{ij}=1) = 2N^{-1}$ otherwise. Thus, the nodes in the same block are more likely to be connected.

\textbf{Example 3.} (Power-Law Distribution Network) It is commonly observed in network analysis that the majority of nodes have few links but a small proportion have a large number of edges \citep{barabasi1999emergence}. Therefore, we simulate the adjacency matrix $A$ according to \cite{clauset2009power}. The in-degree $m_i = \sum_j a_{ji}$ for node $i$ is generated by the discrete power-law distribution with $P(m_i = k) = ck^{-\alpha}$, with a normalizing constant $c$ and $\alpha = 3$. For the $i$th node, $m_i$ nodes are randomly selected to be its followers.

\textbf{Data Generation}. For each node, we generate the true exogenous covariates $\mathbf{x}_i = (x_{i1}, x_{i2})^\top \in \mR^2$ from a multivariate normal distribution with mean $\mathbf{0}_2$ and $\Sigma_x= I_{2}$, and then truncate $\mathbf{x}_i$ by bounded norm $\max_i\|\mathbf{x}_{i}\|_\infty \leq 2$. The corresponding network autoregression coefficient is fixed to be $\rho_0 = 0.2$ and coefficient $\beta_0=(0.3,0.3)^\top$. The response $\mY^*$ is generated based on $\mY^*= (I_N-\rho_0 W)^{-1}(\mX\beta_0+\mE)+\mcE$. Define $X_j=(x_{1j},\cdots,x_{nj})^\top$ for $1\leq j\leq 2$. To simulate privacy-protected covariates, we fix $\mX_1=X_1\in \mR^{N}$ and assume that $\mX_2=X_2\in \mR^{N}$ cannot be faithfully observed. Then, $\mX_2^*$ can be generated by $\mX_2^*=\mX_2+\mcE$ with $\mcE=(\vare_{x,i})$, which follows the setting of the noise distribution.

\textbf{Noise Distribution}. We consider that $\vare_{x,i}$ follows a normal distribution with mean $0$ and $\lambda_x^2=0.5$. We consider three different distributions to generate $\vare_i$ independently with mean $0$ and $\lambda^2=0.5$: (1) a normal distribution, (2) a $t$-distribution with degree 6 and (3) a laplace distribution with location parameter 0 and scale parameter 0.5. It is notable that for the $t$-distribution, each element is divided by $\sqrt{3}$ to make $\var(\vare_i)=\lambda^2=0.5$. For the noise $e_i$, we consider the same cases for $N(0,1)$ and $t(6)$. Here, we set the sample size as $N=$500, 1,000, and 2,000 and consider all the network models.

\textbf{Privacy-Protection Level}. To better illustrate the effect of the privacy-protection level, we consider a fixed sample size $N=1,000$ and the dyad independent network-generation model as an example. Different $\vare_i$s and $\vare_{x,i}$s are generated from normal distributions with variances of $\lambda^2$ and $\lambda^2_x$, respectively. We consider the impact of the variations in $\lambda_x^2$ and $\lambda^2$ on the estimation results. Specifically, for fixed $\lambda^2=0.5$, we consider $\lambda^2_x=(0.2,0.5,0.8)$. For fixed $\lambda^2_x=0.5$, we consider $\lambda^2=(0.2,0.5,0.8).$

\subsection{Performance measurements and simulation results}
For a better comparison, we focus here on the estimation results of $\rho_0$ and $\beta_0$. To gauge the finite-sample performance, we use the following metrics. Define $\hat D^{(r)} = \{\hat d_{j}^{(r)}\}_{j=1}^q \in \mR^q$ as the estimator from the $r$th replication with $q=p+1$. For any $1 \leq j \leq q$, the bias can be evaluated as ${\rm Bias}_j=|\bar{d_{j}}-d_j|$, where $\bar{d_{j}}=R^{-1}\sum_{r}\hat d_{j}^{(r)}$, and $d_j$ is the $j$th element of the true parameter. The standard error can be estimated using $\SE_j=R^{-1}\sum_{r} \SE_j^{(r)}$. Notably, $\widehat{\mbox{SE}}_{j}^{(r)}$ represents the $j$th diagonal element of the estimated asymptotic covariance matrix, which can be computed using (\ref{eq:CLE_Cov}) or (\ref{eq:CLS_Cov}) with $\hat D^{(r)}$ plugged in, followed by a correction step, as discussed in Section 3.3 for the CLE and Section 3.4 for the CLS. Define ${\rm SE}_j=\{R^{-1}\sum_{r}(\hat d_{j}^{(r)}-\bar{d_{j}})^2\}^{1/2}$ as the Monte Carlo standard deviation of $\hat d_{j}^{(r)}$, and the estimation efficiency of $\wh {\rm SE}_j$ can be evaluated by comparing $\SE_j$ and ${\rm SE}_j$. In addition, for each $\hat d_{j}^{(r)}$, a 95\% confidence interval can be constructed as $\mbox{CI}_{j}^{(r)} = [\hat d_{j}^{(r)}-z_{0.975}N^{-1}\widehat{\mbox{SE}}_{j}^{(r)},\hat d_{j}^{(r)}+z_{0.975}N^{-1}\widehat{\mbox{SE}}_{j}^{(r)}]$, where $z_\alpha$ is the lower $\alpha$th quantile of the standard normal distribution. The empirical coverage probability is then evaluated as $\mbox{CP}_{j} = R^{-1}\sum_{r = 1}^R I(\hat d_{j}^{(r)}\in\mbox{CI}_{j}^{(r)} )$, where $I(\cdot)$ is the indicator function. 

\begin{figure}[t]
\subfigure[]{
\includegraphics[width=3.6cm,height=4cm]{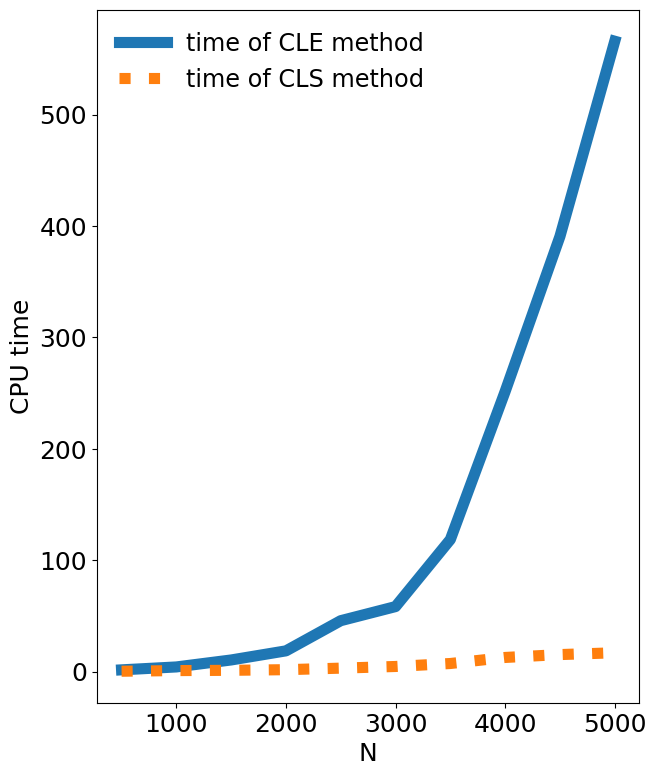} \label{time_Dyad}
}
\hspace{2mm}
\subfigure[]{
\includegraphics[width=3.6cm,height=4cm]{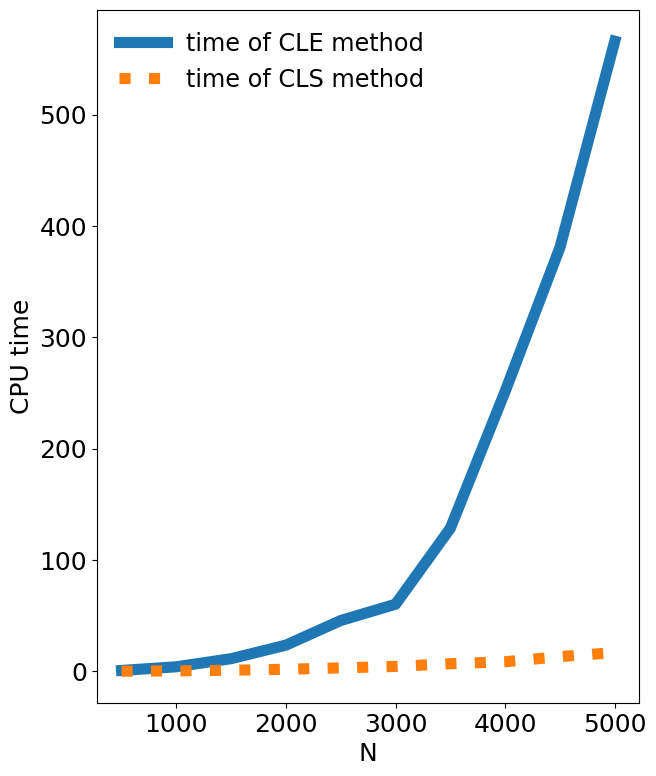} \label{time_SBM}
}	
\hspace{2mm}
\subfigure[]{
\includegraphics[width=3.6cm,height=4cm]{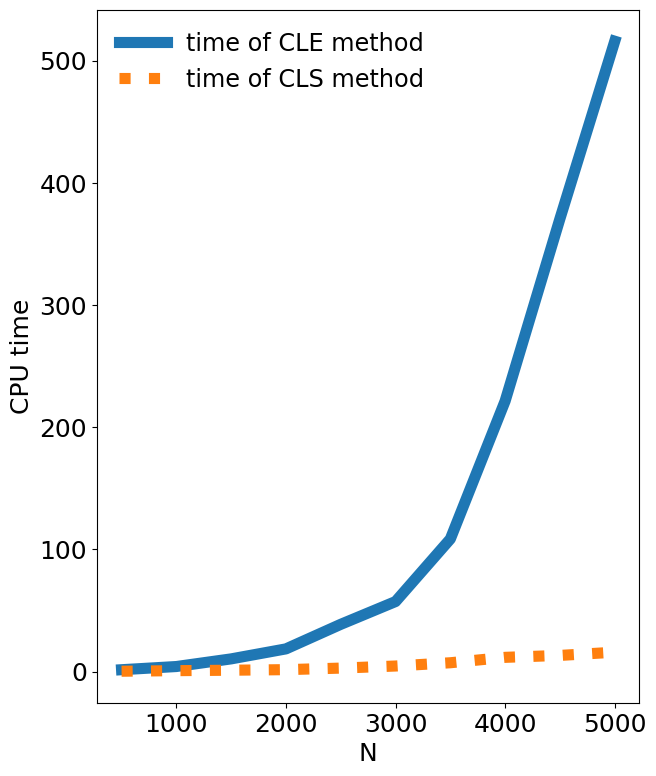} \label{time_Powerlaw}
}

\caption{average CPU time for three simulation examples with 100 replicates under the assumption of a normal noise distribution: Example 1 (left panel), Example 2 (middle panel), and Example 3 (right panel). The solid line is the average CPU time for CLE and the  dashed line is for CLS.
}
\label{time}
\end{figure}

Each experiment is replicated 500 times ($R = 500$). The initial estimators for all the parameters are set to  0.01. All simulations are conducted on a Linux server with a 3.60 GHz Intel Core i7-9700K CPU and 16 GB RAM. 
Table \ref{tab:lambda} presents the estimation outcomes for various privacy-protection levels under dyad independence model with $N=1000$, where $\mE$ and $\mcE$ follow a normal distribution. The evaluation metrics include Bias, SE, and CP across different values of $\lambda^2$ and $\lambda_x^2$. Figure \ref{time} displays the average CPU time (in seconds)  required for CLE and CLS, facilitating a comparison of computational efficiency. Due to space constraints, additional estimation results for varying sample sizes, network structures, and noise distributions are provided in Tables E.1–E.5 in the Appendix.

\textbf{Estimation Performance.} From Table \ref{tab:lambda} and Tables E.1–E.5, we can see that the classical QMLE is seriously biased as expected. Thus we focus on the performance of the proposed estimator CLE and CLS. We draw the following conclusions. First, the estimation bias is sufficiently small for all sample sizes and both the methods. Second, as the sample size $N$ increases, Bias, SE, and $\SE$ all decrease, which shows the consistency of both methods. Third, all the CP values are approximately 95\% for $\alpha=0.05$ with $\SE$ and SE clearly being close to each other. This corroborates the theoretical conclusions in Theorem \ref{thm:cle} for the CLE and in Theorem \ref{thm:CLS} for the CLS. Finally, CLE has a smaller SE than CLS.

\textbf{Effect of Privacy-Protection Level.} From Table \ref{tab:lambda}, we can see the bias of $\rho$ or $\beta$ of the classical QMLE increases as the noise variance $\lambda^2$ or $\lambda_x^2$ increases, respectively. Based on the results of CLE and CLS in Table \ref{tab:lambda}, we can conclude that an increase in the privacy-protection level leads to a larger $\widehat{\rm SE}$ for the estimators. Specifically, when $\lambda_x^2$ is fixed, increasing $\lambda^2$ leads to an increase in $\widehat{\rm SE}$ for all estimators. However, when $\lambda^2$ is fixed, as $\lambda_x^2$ increases, only $\widehat{\rm SE}$ for the estimator corresponding to the unobserved $\mX_2$ increases, whereas those for the other estimators remain relatively stable.

\textbf{Computation Time.} We conduct the same experiment as previously, but with a normal noise distribution and fixed $\sigma_0^2=1,$ $\lambda_x^2=1,$ and $\lambda^2=0.5$. Moreover, we change $N$ from 500 to 5,000 for $R = 100$ (100 replicates) to further compare the computational efficiency of CLE and CLS. Figure \ref{time} shows the results from a computational perspective. Evidently, as the sample size $N$ increases, the computational time of CLE increases dramatically, whereas that of CLS increases much more slowly. In practice, a method should be chosen based on the balance between computational time and statistical accuracy.

\section{ Restaurant Data Analysis}
We analyze  the restaurant data from small- and medium-sized restaurants provided by a leading third party payment platform in China.
 The dataset comprises aggregated transaction records for fast-food restaurants and their customers in Guangzhou, China, covering the period from January 1 to July 1, 2024; all variables have been pre-processed by the payment platform into time-aggregated measurements. 
To explore the impact of other restaurants on the performance of a specific restaurant, we first develop a PSAR model for restaurant network analysis. Next, we examine the interaction between restaurant and customer responses using the Privacy-Protected MMSAR model. The descriptive analysis was provided by the platform, and the non-protected SAR benchmark was obtained by the platform using our code on the raw data, while the PSAR analysis was conducted on the noise-added data released to us.

\subsection{Restaurant Network Analysis based on PSAR \label{Restaurant Analysis}} This dataset contains information from \(N = 1996\) fast-food restaurants. For each restaurant \(i\) \((1 \leq i \leq N)\), the response \(Y_i\) is defined as the daily transaction volume of the restaurant during the observation period. Additionally, we consider five covariates for each restaurant: (1) \textit{average transaction value} \(X_1\), the average price per transaction at the restaurant; 
(2) \textit{repeat customer ratio} \(X_2\), representing the percentage of customers who dined at the restaurant two or more times in the previous two months; (3) \textit{operating hours} \(X_3\), where \(x_{3i} = 1\) indicates that the restaurant operates during both lunch and dinner hours, and \(x_{3i} = 0\) otherwise; 
(4) \textit{commercial zone diversity} \(X_4\), measured by the number of types of commercial zones in the vicinity of a restaurant’s location (e.g., office areas, residential neighborhoods, hotels, schools, hospitals, dining districts, shopping malls, and tourist attractions);
(5) \textit{menu variety} \(X_5\), measured by the number of distinct dishes offered by the restaurant.

All the continuous variables are standardized. 
For privacy protection, the platform added Laplace noise with variance $0.2$ to the response $Y$ and to all covariates except $X_3$, which is publicly available without modification,
which means $\lambda_x^2=\lambda^2=0.2$.
To analyze the effect of the network structure, if two restaurants \(i\) and \(j\) have at least five shared customers, an edge is defined with \(a_{ij}=1\); otherwise, \(a_{ij}=0\).
Furthermore, the network density (i.e., \(\{N(N-1)\}^{-1}\sum_{i_1,i_2}a_{i_1i_2}\)) is 0.31\%, which implies that this is a relatively sparse network 
and suggests that
localized competition may exist. To further protect the privacy of the network structure, the platform randomly perturbs \(5\log(N)\) edges in the network.

\begin{table}[t]\footnotesize
\caption{Non-protected benchmark results (NSAR) for restaurant network analysis and CLE and CLS results using privacy-protected SAR. For each estimator, “*” indicates significance at the 0.05 level.}
\label{tab:real_competition}
\begin{center}
\begin{tabular}{cccccc}
\hline
\hline
&NSAR(SE)&CLE(SE)&CLS(SE)\\
\hline
Intercept  &0.033(0.018) &0.023(0.023)  & 0.044(0.026)  \\
Network Effect  & -0.133(0.018)*  &-0.120(0.025)*&-0.145(0.029)*   \\
Average Transaction Value& 0.215(0.018)* & 0.226(0.024)* &0.221(0.028)* \\
Repeat Customer Ratio  & 0.371(0.018)*&0.331(0.023)*& 0.362(0.026)* \\
Operating Hours    &0.071(0.018)* &0.069(0.024)*& 0.058(0.027)*  \\
Commercial Zone Diversity & 0.089(0.018)* & 0.093(0.023)* & 0.085(0.028)*\\
Menu Variety & 0.019(0.018) & 0.012(0.023) & 0.023(0.028)\\
\hline
RMSE         &  1.129  &1.335  &1.820    \\
Time (seconds)         &  30.416  &50.901  &8.812  \\
\hline
\hline
\end{tabular}
\end{center}
\end{table}

Then, we estimate the model using the CLE and CLS estimators and also the non-protected benchmark estimators (NSAR). In addition to reporting the measurements from the simulation results, we also provide the root mean square error (RMSE) of each method for comparison. To be more precise, define $(\hat\rho,\hat\beta)$ as the estimators of $(\rho_0,\beta_0)$ (i.e., NSAR, CLE or CLS). The RMSE is calculated by $ \mbox{RMSE} =\{N^{-1}\sum_{i=1}^N(\hat{Y_i}-Y_i)^2\}^{1/2}$, and $ (\hat{Y_1},\dots,\hat{Y}_N)^T=(I_N-\hat\rho W)^{-1}\mX^{*}\hat\beta$.
The estimation results are summarized in Table \ref{tab:real_competition}. 

The analysis reveals that transaction volumes are positively associated with higher average prices and repeat customer ratios, underscoring the importance of pricing strategies and customer loyalty. Operating during both lunch and dinner hours and being located in diverse commercial areas are linked to better transaction performance, whereas menu variety shows no significant effect. After accounting for these factors, the network effect remains statistically significant and negative, indicating a competitive relationship. Both methods (CLE \& CLS) provide similar estimates for network effects and covariates with the NSAR estimator, showing that privacy-protected SAR models with perturbed network maintain robust results. In terms of model performance, CLS shows a slightly larger RMSE than CLE, suggesting CLE offers better predictive accuracy. Besides, CLS requires about one-sixth of the computational time of CLE, as it avoids calculating $\Omega^{-1}$.

\subsection{Restaurant-Customer Network Analysis based on Privacy-Protected MMSAR}
Interactions arise when customers dine frequently at the same restaurant. To model this, we define a two-mode network where connections are formed between restaurants and customers, with no direct links existing among restaurants or among customers themselves. Specifically, an undirected edge is established $(a_{i_1i_2} = 1)$ between restaurant $i_1$ $(1 \leq i_1 \leq N_1)$ and customer $i_2$ $(1 \leq i_2 \leq N_2)$ if customer $i_2$ has dined at restaurant $i_1$ more than five times. This threshold-based network construction serves to identify meaningful, sustained patronage relationships while maintaining computational tractability, resulting in a sparse network.
Given the computational complexity of MMSAR, the platform randomly sampled the network, resulting in $N_1 = 833$ restaurants and $N_2 = 1660$ customers. The network density is 0.17\%.
Customers are unaware of a restaurant's total transaction volume but are aware of the transaction amount per customer and how busy the restaurant is.
Correspondingly, two response variables for restaurants are defined as: (1) \textit{average transaction value} $Y_{11}$, the average amount per transaction; (2) \textit{daily transaction frequency} $Y_{12}$, the average number of transactions per day. The customers' response variable is \textit{daily spending amount}, denoted as $Y_2$.
The restaurant-level covariates in this analysis consist of the significant predictors identified in the transaction volume analysis in Section \ref{Restaurant Analysis}: \textit{repeat customer ratio} $X_{12}$, \textit{operating hours} $X_{13}$, and \textit{commercial zone diversity} $X_{14}$.
For customer covariates, we include: (1) \textit{proportion of lunch and dinner} $X_{21}$, the share of daily spending during lunch and dinner; (2) \textit{coupon usage} $X_{22}$, with 0 indicating no coupon usage and 1 indicating at least one instance of coupon use. All continuous variables are standardized. For privacy protection, 
the platform has added Laplace noise with a variance of $0.2$ to all continuous covariates and responses.

We estimate the model using the NSAR, CLE and CLS methods and present the results in Table \ref{realdata_inter}.
From these results, we draw the following insights about the dataset. First, daily customer spending ($Y_2$) positively affects average price ($Y_{11}$). However, the reverse effect is not evident: a higher average price does not necessarily attract more high-spending customers. Meanwhile, daily transaction frequency ($Y_{12}$) negatively impacts customer spending ($Y_{2}$), suggesting that busier restaurants may compromise service quality or dining experience, leading to lower per-customer expenditure. This implies that merchants should focus on enhancing service quality and the dining experience to retain high-spending customers, rather than relying solely on price increases. Additionally, managing customer flow to avoid overcrowding can help maintain per-customer spending and overall transaction value.
 Restaurant covariates exhibit patterns similar to those in Section  \ref{Restaurant Analysis}: repeat customer ratio $X_{12}$ and extended operating hours $X_{13}$ significantly boost performance, while commercial zone diversity $X_{14}$ underscores the advantage of diversified commercial surroundings. On the customer side, spending during peak meal times $X_{21}$ is positively related to higher daily customer spending, whereas coupon usage $X_{22}$ shows no significant effect.  
In terms of model performance, the estimates among NSAR, CLE and CLS methods are similar within twice the standard error, which demonstrates the robustness of the privacy-protected MMSAR framework, even with network perturbation and noise injection. Notably, CLS maintains computational efficiency, validating its utility for large-scale analyses.

\begin{table}[H]\footnotesize
\caption{Non-protected benchmark results (NSAR) for restaurant-customer network analysis and CLE and CLS results using privacy-protected MMSAR. For each estimator (i.e., CLE and CLS), ``*'' and ``**'' indicate significance at the 0.05 and 0.01 levels, respectively.}
\label{realdata_inter}
\begin{center}
\begin{tabular}{c|cc|cc|cc}
\toprule
\toprule
&\multicolumn{2}{c|}{NSAR(SE)}&\multicolumn{2}{c|}{CLE(SE)}&\multicolumn{2}{c}{CLS(SE)}\\
\midrule
 Customer to Restaurant&$Y_{11}$&$Y_{12}$&$Y_{11}$&$Y_{12}$&$Y_{11}$&$Y_{12}$\\
\hline
Daily Spending Amount($Y_2$)&\makecell[c]{0.239*\\(0.041)}&\makecell[c]{0.003\\(0.011)}&\makecell[c]{0.259*\\(0.048)} &\makecell[c]{0.003\\(0.013)} &\makecell[c]{0.280*\\(0.055)} &\makecell[c]{0.004\\(0.016)} \\
\hline
Repeat Customer Ratio($X_{12}$) &\makecell[c]{0.211*\\(0.040)}&\makecell[c]{0.438*\\(0.042)}   &\makecell[c]{0.258*\\(0.048)} &\makecell[c]{0.392*\\(0.050)}  &\makecell[c]{0.282*\\(0.061)}&\makecell[c]{0.364*\\(0.055)}   \\
Operating Hours($X_{13}$) &\makecell[c]{0.189*\\(0.041)}&\makecell[c]{0.101*\\(0.041)} &\makecell[c]{0.175*\\(0.045)} &\makecell[c]{0.097*\\(0.045)} &\makecell[c]{0.194*\\(0.053)}&\makecell[c]{0.125*\\(0.054)} \\
Commercial Zone Diversity($X_{14}$)&\makecell[c]{0.271*\\(0.039)}&\makecell[c]{0.385*\\(0.040)} &\makecell[c]{0.253*\\(0.044)}&\makecell[c]{0.450*\\(0.045)}&\makecell[c]{0.280*\\(0.052)}&\makecell[c]{0.489*\\(0.049)}\\
\hline
RMSE&1.024 &1.142 &1.163&1.623&1.349&2.241\\
\midrule[1pt]
Restaurant to Customer& \multicolumn{2}{c|}{$Y_2$}& \multicolumn{2}{c|}{$Y_2$}&\multicolumn{2}{c}{$Y_2$}\\
\hline
  Average Transaction Value($Y_{11}$)&\multicolumn{2}{c|}{\makecell[c]{-0.031\\(0.040)}}  &\multicolumn{2}{c|}{\makecell[c]{-0.042\\(0.045)}} & \multicolumn{2}{c}{\makecell[c]{-0.027\\(0.052)}} \\
  Daily Transaction Frequency($Y_{12}$)&\multicolumn{2}{c|}{\makecell[c]{-0.125*\\(0.030)}} &\multicolumn{2}{c|}{\makecell[c]{-0.116*\\(0.031)}}  &\multicolumn{2}{c}{\makecell[c]{-0.092*\\(0.036)}}  \\
  \hline
  Proportion of Lunch and Dinner ($X_{21}$)&\multicolumn{2}{c|}{\makecell[c]{0.163**\\(0.031)}} &\multicolumn{2}{c|}{\makecell[c]{0.182*\\(0.036)}} &\multicolumn{2}{c}{\makecell[c]{0.136*\\(0.040)}} \\
  Coupon Usage($X_{22}$)&\multicolumn{2}{c|}{\makecell[c]{-0.027\\(0.029)}} &\multicolumn{2}{c|}{\makecell[c]{-0.056\\(0.035)}} &\multicolumn{2}{c}{\makecell[c]{-0.043\\(0.045)}}\\
  \hline
  RMSE &\multicolumn{2}{c|}{1.196}&\multicolumn{2}{c|}{1.421}&\multicolumn{2}{c}{1.927}\\
\midrule
 Time (seconds)&\multicolumn{2}{c|}{342.85} &\multicolumn{2}{c|}{782.28}&\multicolumn{2}{c}{59.31}\\
\bottomrule
\bottomrule
\end{tabular}
\end{center}
\end{table}

\section{Concluding remarks}

This study addressed the issue of privacy protection in the SAR model with noise-added response and covariates. We focused on achieving a consistent estimation of the model parameters and proposed two different estimators: the CLE and the CLS. Both of the estimators provide reliable solutions in privacy-protected scenarios. Theoretical properties have been carefully established and model generalizations are discussed.

For future research, three directions are proposed. First, while we focused on common SAR models and their extensions, studying estimators in dynamic and nonlinear networks with added noise remains essential. Second, Condition (C5) imposes relatively strict requirements; relaxing these conditions to enhance network edge privacy requires further exploration. 
Finally, from the application perspective, incorporating richer interpretable covariates, such as restaurant size and seating capacity, would be valuable; from the theoretical perspective, investigating privacy-protected spatial models with high-dimensional features is also a meaningful direction.

\par\bigskip
\noindent\textbf{Acknowledgments.} The authors would like to thank the Editor, the Associate Editor, and the referees for their constructive comments and advice that improved the quality of this paper. Shuyuan Wu is the correpsonding author. Danyang Huang's research is supported by the NSFC (72471230), the National Key R\&D Program of China (2023YFC3304701), the MOE Project of Key Research Institute of Humanities and Social Sciences (22JJD110001), Chinese institute for public governance research, Big Data and Responsible Artificial Intelligence for National Governance, Renmin University of China. Shuyuan Wu's research is partially supported by the National Natural Science Foundation of China (12401392). Hansheng Wang's research is partially supported by National Natural Science Foundation of China (72495123, 12271012).

\par\bigskip
\noindent\textbf{Supplementary material.} The supplementary material follows the main article and contains necessary notation and lemmas, proofs of the main theorems, and additional simulation results.


\bibliographystyle{imsart-nameyear} 
\bibliography{ref}       





\begin{table}[H]\footnotesize
{\caption{Simulation results with 500 replicates for different privacy-protection levels under dyad independence model with $N=1000$ and $\mE$ and $\mcE$ following a normal distribution. The bias, SE, $\SE$, and CP are reported for QMLE, CLE and CLS.}\label{tab:lambda}
\begin{center}
\begin{tabular}{ccc|ccc|ccc|ccc}
\hline
\hline
& & &\multicolumn{3}{c}{QMLE}&\multicolumn{3}{c}{CLE} &\multicolumn{3}{c}{CLS}\\
{$\lambda^2_x$}&{$\lambda^2$} &{Est.} & $\rho$&$\beta_1$&$\beta_2$ & $\rho$&$\beta_1$&$\beta_2$ & $\rho$&$\beta_1$&$\beta_2$ \\
\hline
0.2&0.5 & Bias &	0.060&0.000&0.043&	0.002	&	0.000	&	0.002	&	0.001	&	0.000	&	0.002\\
& & SE &0.051&0.036&0.029 &	0.040	&	0.035	&	0.041	&	0.072	&	0.040	&	0.042	\\
& & $\SE$&0.052&0.037&0.029&	0.042	&	0.036	&	0.042	&	0.071	&	0.039	&	0.041	\\
& & CP&72.8&94.6&82.2&	96.6	&	94.6	&	96.4	&	94.0	&	93.4	&	94.2	\\
&&&&&&&&& \\ [-1.1em]
0.5&0.5 & Bias&0.062&0.001&0.100&0.002	& 0.000	&	0.002	&	0.002	&	0.001	&	0.002	\\
& & SE&0.054&0.039&0.031&0.043 &	0.039	&	0.048	&	0.083	&	0.042	&	0.053	\\
& & $\SE$&0.052&0.039&0.032&0.044	&	0.039	&	0.049	&	0.081	&	0.043	&	0.058	\\
& & CP&79.0&94.0&9.2&94.8	&	95.0	&	95.8	&	95.2	&	94.8	&	96.2	\\
&&&&&&&&& \\ [-1.1em]											
0.8&0.5 & Bias&0.061&0.001&0.128&	0.002	&	0.001	&	0.001	&	0.002	&	0.001	&	0.002	\\
& & SE&0.056&0.041&0.039&	0.076	&	0.043	&	0.055	&	0.085	&	0.043	&	0.062	\\
& & $\SE$&0.055&0.043&0.038&	0.075	&	0.044	&	0.054	&	0.086	&	0.044	&	0.065	\\
& & CP&75.2&95.0&13.8&	94.8	&	95.4	&	95.4	&	93.2	&	94.2	&	96.6	\\
\hline					
0.5&0.2& Bias&0.031&0.001&0.102&	0.004	&	0.000	&	0.002	&	0.003	&	0.000	&	0.002	\\
& & SE&0.050&0.047&0.043&	0.040	&	0.036	&	0.045	&	0.069	&	0.039	&	0.052	\\
& & $\SE$&0.049&0.049&0.040&	0.041	&	0.035	&	0.046	&	0.064	&	0.040	&	0.053	\\
& & CP&89.2&95.2&28.6&	95.6	&	95.4	&	96.4	&	94	&	95.2	&	96.4	\\
&&&&&&&&& \\ [-1.1em]											
0.5&0.5 & Bias&0.062&0.001&0.100&0.002	& 0.000	&	0.002	&	0.002	&	0.001	&	0.002	\\
& & SE&0.054&0.039&0.031&0.043 &	0.039	&	0.048	&	0.083	&	0.042	&	0.053	\\
& & $\SE$&0.052&0.039&0.032&0.044	&	0.039	&	0.049	&	0.081	&	0.043	&	0.058	\\
& & CP&79.0&94.0&9.2&94.8	&	95.0	&	95.8	&	95.2	&	94.8	&	96.2	\\
&&&&&&&&& \\ [-1.1em]											
0.5&0.8 & Bias&0.085&0.003&0.099&	0.002	&	0.001	&	0.001	&	0.001	&	0.002	&	0.002	\\
& & SE&0.048&0.039&0.035&	0.052	&	0.045	&	0.056	&	0.097	&	0.043	&	0.063	\\
& & $\SE$&0.049&0.041&0.036&	0.057	&	0.047	&	0.053	&	0.096	&	0.044	&	0.064	\\
& & CP&58.8&92.6&46.4&	95.0	&	95.4	&	93.4	&	94.6	&	95.8	&	96.2	\\
\hline
\hline
\end{tabular}
\end{center}}
\end{table}

\end{document}